\documentclass[manuscript,screen,nonacm]{acmart}
\usepackage{color, colortbl, xcolor}
\usepackage{url}
\usepackage{subcaption}
\usepackage{textcomp}
\usepackage{soul}
\usepackage{multirow}
\usepackage{enumitem}
\usepackage{mathtools}
\usepackage{siunitx}
\usepackage{array}
\usepackage{colortbl}
\usepackage{hhline}
\usepackage[table]{xcolor}
\usepackage{booktabs}

\newcommand{\distbar}[1]{\rule[0pt]{3.5pt}{#1}\hspace{1.5pt}}
\newcommand{\dist}[5]{\distbar{#1}\distbar{#2}\distbar{#3}\distbar{#4}\distbar{#5}}
\usepackage{booktabs} % For formal tables | Already in ACM
\usepackage{array}
\usepackage{xcolor}

\newcommand*{\rowstyle}[1]{% sets the style of the next row
  \gdef\@rowstyle{#1}%
  \@rowstyle\ignorespaces%
}

\newcolumntype{=}{% resets the row style
  >{\gdef\@rowstyle{}}%
}

\newcolumntype{+}{% adds the current row style to the next column
  >{\@rowstyle}%
}

\usepackage{arydshln}
\definecolor{linkColor}{RGB}{6,125,233}
\definecolor{green}{rgb}{0.0, 0.65, 0.31}
\definecolor{bleudefrance}{rgb}{0.19, 0.55, 0.91}
\definecolor{ceruleanblue}{rgb}{0.16, 0.32, 0.75}
\definecolor{grey}{HTML}{969696}
\definecolor{violet}{HTML}{756bb1}
\definecolor{dgrey}{HTML}{01665e}
\definecolor{lgrey}{HTML}{5ab4ac}
\definecolor{dgreen}{HTML}{005a32}
\definecolor{purple}{HTML}{54278f}

\definecolor{editCol}{HTML}{000000}
\definecolor{maskCol}{HTML}{c51b7d}
\definecolor{lrColor}{HTML}{8856a7}
\definecolor{trColor}{HTML}{d01c8b}
\definecolor{ctColor}{HTML}{4dac26}
\definecolor{brickred}{HTML}{f03b20}
\definecolor{improveCol}{HTML}{253494}
\definecolor{worsenCol}{HTML}{d7191c}
\definecolor{DarkBlue}{HTML}{00008B}
\definecolor{mscolor}{HTML}{01665e}
\definecolor{nmscolor}{HTML}{bf812d}
\definecolor{lgreen}{HTML}{ccece6}
\definecolor{dolive}{HTML}{308014}
\definecolor{dpink}{HTML}{CD1076}
\definecolor{soothinggreen}{HTML}{4dac26}
\definecolor{darkred}{HTML}{8B0000}

\colorlet{tablerowcolor4}{gray!50} % Table row separator colour = 

\newcommand*{\textlabel}[2]{%
  \edef\@currentlabel{#1}% Set target label
  \phantomsection% Correct hyper reference link
  #1\label{#2}% Print and store label
}

\colorlet{tableheadcolor}{gray!25} % Table header colour = 25% gray
\colorlet{tablerowcolor}{gray!10} % Table row separator colour = 
\colorlet{tablerowcolor2}{gray!45} % Table row separator colour = 
\colorlet{tablerowcolor3}{gray!25} % Table row separator colour = 10% gray

\newcommand{\rowcollight}{\rowcolor{tablerowcolor3}} %

\newcolumntype{a}{>{\columncolor{tablerowcolor}}r}
\definecolor{aicolor}{HTML}{018571}
\definecolor{occolor}{HTML}{ff7799}

\definecolor{aicolor}{HTML}{fc8d62}
\definecolor{occolor}{HTML}{253494}

\newif{\ifhidecomments}
 \hidecommentsfalse 
\ifhidecomments
    \newcommand{\jenny}[1]{}
    \newcommand{\dongwhi}[1]{}
    \newcommand{\eshwar}[1]{}
    \newcommand{\koustuv}[1]{}
\else
    \newcommand{\jenny}[1]{\textbf{\small\sffamily{\textcolor{DarkBlue}{[#1 -- Jenny]}}}}
     \newcommand{\dongwhi}[1]{\textbf{\small\sffamily{\textcolor{dolive}{[#1 -- Dong Whi]}}}}
    \newcommand{\eshwar}[1]{\textbf{\small\sffamily{\textcolor{brickred}{[#1 -- Eshwar]}}}}
    \newcommand{\koustuv}[1]{\textbf{\small\sffamily{\textcolor{dpink}{[#1 -- Koustuv]}}}}
  \fi

\usepackage[most]{tcolorbox}

\newtcolorbox{takeawaybox}{
    colback=gray!10,
    colframe=gray!35,
    boxrule=0.4pt,
    arc=2pt,
    left=6pt,
    right=6pt,
    top=4pt,
    bottom=4pt,
    before skip=6pt,
    after skip=8pt
}

\renewcommand{\textrightarrow}{$\rightarrow$}

\newcommand{\ppen}{\textsc{PeerPen}}
\colorlet{tableheadcolor}{gray!25} % Table header colour = 25% gray

\definecolor{neutralCol}{HTML}{dd1c77}
\definecolor{neutralGreen}{HTML}{31a354}
\definecolor{NewBlue}{HTML}{1879ba}
\definecolor{bleudefrance}{rgb}{0.19, 0.55, 0.91}  
\definecolor{AfTrColor}{HTML}{0868ac}  
\definecolor{BfTrColor}{HTML}{a8ddb5}  

\definecolor{AfCtColor}{HTML}{b10026}  
\definecolor{BfCtColor}{HTML}{fd8d3c}

\graphicspath{ {figures/} }

\newcommand{\para}[1]{\vspace{0.3em}\noindent\textbf{#1}~}

\newcolumntype{C}[1]{>{\centering\arraybackslash}p{#1}}

\AtBeginDocument{%
  \providecommand\BibTeX{{%
    \normalfont B\kern-0.5em{\scshape i\kern-0.25em b}\kern-0.8em\TeX}}}

\begin{document}


% \title[\llum{}: System Design and Evaluation]{\llum{}: System Design and Evaluation}

\title[\ppen{}: AI-Assisted Writing for Online Mental Health Peer Support]{\ppen{}: AI-Assisted Writing for Online Mental Health Peer Support}


% \renewcommand{\shortauthors}{Koustuv Saha et al.}

\author{Jiwon Kim}
\orcid{0009-0002-5182-060X}
\affiliation{%
  \institution{University of Illinois Urbana-Champaign}
 \city{Urbana}
 \state{IL}
 \country{USA}
}
\email{jiwonk7@illinois.edu}

\author{Sherry Gong}
\orcid{0009-0005-4708-9932}
\affiliation{%
  \institution{University of Illinois Urbana-Champaign}
 \city{Urbana}
 \state{IL}
 \country{USA}
}
\email{sgongg@illinois.edu}

\author{Maya Ajit}
\orcid{0009-0000-2677-9268}
\affiliation{%
  \institution{University of Illinois Urbana-Champaign}
 \city{Urbana}
 \state{IL}
 \country{USA}
}
\email{ajit2@illinois.edu}

\author{Soorya Ram Shimgekar}
\orcid{0000-0002-1110-9699}
\affiliation{%
  \institution{University of Illinois Urbana-Champaign}
 \city{Urbana}
 \state{Illinois}
 \country{USA}
}
\email{sooryas2@illinois.edu}

\author{Yunhao Yuan}
\orcid{0000-0002-1450-8572}
\affiliation{%
  \institution{Aalto University}
 \city{Espoo}
 \state{Illinois}
 \country{Finland}}
\email{yunhao.yuan@aalto.fi}

\author{Dong Whi Yoo}
\orcid{0000-0003-2738-1096}
\affiliation{%
 \institution{Indiana University Indianapolis}
 \city{Indianapolis}
 \state{IN}
 \country{USA}}
 \email{dy22@iu.edu}

\author{Eshwar Chandrasekharan}
\orcid{0000-0002-7473-1418}
\affiliation{%
  \institution{University of Illinois Urbana-Champaign}
 \city{Urbana}
 \state{IL}
 \country{USA}}
\email{eshwar@illinois.edu}

\author{Koustuv Saha}
\orcid{0000-0002-8872-2934}
\affiliation{%
  \institution{University of Illinois Urbana-Champaign}
 \city{Urbana}
 \state{IL}
 \country{USA}}
\email{ksaha2@illinois.edu}

\renewcommand{\shortauthors}{Jiwon Kim et al.}

\begin{abstract}

Online mental health communities thrive on peer support, yet those who volunteer to help often lack formal training and may struggle to articulate supportive responses. AI co-writing could lower this barrier; however, peer support derives much of its value from being perceived as personal, raising questions around authorship, ownership, and trust. We built \ppen{}, a writing assistance tool embedded within a Reddit-like interface, supporting two main features: draft generation and revision of user-written responses. Through semi-structured interviews with 15 participants, we find that \ppen{} reduced the burden of composing responses and increased confidence in offering support. Participants wanted AI to assist their writing without taking over authorship and anticipated tensions around authenticity and trust. Such assistance could make authorship uncertain even for responses written without it, weakening trust across the community. We contribute design implications for AI writing assistance that scaffolds supportive communication, preserves authorship, and accounts for community-level trust.

\end{abstract}





\maketitle

% \ifhidecomments
% \else
%     \thispagestyle{firststyle} % applies firststyle to first page
%     \pagestyle{allstyle}
% \fi

% \input{00notes.tex}
\section{Introduction}

Online communities have become an important source of social support for individuals experiencing psychological distress, loneliness, and other mental health challenges~\cite{de2014mental,saha2020omhc,andalibi2017sensitive,naslund2016future}. 
Platforms such as Reddit mental health communities (e.g., r/depression, r/Anxiety, r/SuicideWatch), TalkLife, and 7Cups enable people to seek and provide mental health support ~\cite{de2016discovering,de2017language,andalibi2016understanding,saha2020omhc,sharma2018mental}. 
Although online mental health communities (OMHCs) offer timely and accessible support, much of it comes from non-expert peers whose responses, however well intentioned, may not always be effective~\cite{saha2020causal}.
Community members often struggle to find appropriate words, balance empathy with practical guidance, and avoid unintentionally causing harm~\cite{althoff2016large,kazdin2011evidence,torous2018empowering}, and many choose not to respond at all despite wanting to help, limiting the availability of support within these communities.

Recent advances in large language models (LLMs) have shown promise in generating supportive and empathetic responses through conversational agents, AI counselors, and supportive response generation systems that can assist individuals during emotionally sensitive interactions~\cite{sharma2023cognitive, dasswain2025ai, saha2025ai, li2025counselbench, nguyen2026calm}. 
These studies have largely focused on the quality, safety, and empathy of AI-generated responses, and less is known about how people collaborate with AI when supporting others.
In practice, supportive communication is unlikely to become fully automated; rather, AI may assist people as they formulate and refine what they want to say. 
Yet, how such assistance is perceived, used, and negotiated by peer supporters remains underexplored.

Further, AI writing assistance is increasingly becoming part of interpersonal communication, with people using AI to modify, generate, and refine messages they send to others~\cite{hancock2020ai,meng2025ai}, and peer supporters may similarly turn to general-purpose assistants when composing support~\cite{meng2025ai}. 
However, doing so may require transferring sensitive community content to external systems, raising questions about how such data are processed, stored, and governed~\cite{zhong2025considerations}. 
Integrating AI assistance directly within OMHCs presents a unique design opportunity, allowing sensitive data and AI assistance to be governed in ways that better align with existing community practices and norms~\cite{peng2020exploring,wang2025practice,kim2026llumi}, making it an important direction to explore.

Therefore, AI writing assistance should not only produce high-quality responses, but also fit naturally into peer supporters' writing processes and maintain the trust and social norms shaping the experiences of both support seekers and providers~\cite{de2014mental}.
Designing AI that responsibly augments supportive interactions thus calls for examining assistance as part of the peer-support writing process, rather than as a response-generation and model-tuning problem alone.
Accordingly, we ask the research question (RQ): \textbf{How do peer supporters perceive and engage with AI-assisted writing for online mental health support, and what opportunities and challenges does this reveal for design?} 
% To bridge the above gap, our study addresses two research questions (RQs):

Towards our RQ, we derive a set of theory-driven design goals, which guide the design of \ppen{}, an AI writing assistance tool on a Reddit-like interface.
\ppen{} provides two complementary forms of assistance: a \textit{Generation Feature} (GF) that drafts an initial supportive response, and an \textit{Improvement Feature} (IF) that revises user-written responses to enhance empathy, supportiveness, and emotional appropriateness while preserving the user's original intent. 
This contrast between AI-generated drafts and localized revisions of user-written text allows us to examine how different forms of AI assistance shape users' sense of authorship.
By positioning AI as a collaborative writing assistant, \ppen{} aims to reduce the effort of providing support while preserving users' agency over what they ultimately choose to share.
Next, we conduct a user study with 15 participants who use \ppen{} to respond to real posts from OMHCs and reflect on their experiences of writing a response through semi-structured interviews. 

We analyze the interview transcripts using reflexive thematic analysis~\cite{braun2019reflecting}.
Participants saw value in \ppen{} for reducing the effort of composing responses, offering actionable suggestions, and helping them participate in supportive conversations, and described how revision surfaced alternative ways of communicating support.
However, they also emphasized that AI-assisted writing should remain natural, contextually appropriate, and aligned with their own writing style. 
Participants wanted AI to assist their writing without taking over authorship, so that they could decide what to keep, change, and ultimately send.
Additionally, they raised concerns about how AI assistance could affect trust among peer supporters, support seekers, and the broader online community. 
Importantly, we identify \textbf{authorship spillover}: participants anticipated that, if AI writing assistance became widely available, uncertainty about AI involvement could extend beyond AI-assisted responses themselves, making even entirely human-written responses subject to doubt.

The contributions of our work are three-fold. 
First, we make a \textbf{systems contribution}: we design and build \ppen{}, an AI-assisted writing tool that embeds assistance at two points in the peer-support writing process---AI-initiated drafting and localized revision of user-authored text---
and evaluate it with participants responding to real posts from OMHCs. 
Second, we provide an \textbf{empirical account} of process-dependent authorship in AI-assisted peer support. Participants' sense of ownership depended not only on how much text the AI supplied, but on whether they remained the source of communicative intent and retained decision authority over how that intent was expressed.
Third, we derive \textbf{design recommendations} for AI-assisted peer-support writing: scaffolding rather than replacing supportive communication, in-situ guidance that preserves authorship, and attention to transparency and community norms.
Together, these contributions show that AI-assisted peer support must be understood not only through what the system writes, but also how people write with it and how the availability of such assistance may reshape expectations and trust within the broader community. %1.25 pages
\section{Related Work}

\subsection{Peer Support in Online Mental Health Communities}

Online mental health communities (OMHCs) serve as venues for peer-based social support, where members share personal experiences and receive empathy, encouragement, and advice from others with similar experiences~\cite{de2014mental,wadden2021effect,vornholt2021understanding,kim2023supporters,sharma2018mental}. 
They can be particularly valuable when formal care is inaccessible, delayed, or stigmatized, and when people are reluctant to discuss sensitive experiences in identifiable or face-to-face settings~\cite{de2014mental,andalibi2016understanding,vornholt2021understanding}; anonymity and reduced identifiability facilitate candid self-disclosure and help-seeking~\cite{de2014mental,andalibi2016understanding}. Such exchanges can improve emotional wellbeing and foster belonging, though support quality varies and responses can be harmful or dismissive~\cite{de2014mental,saha2020causal,wadden2021effect}.

However, the presence of peer support does not ensure that responses are effective~\cite{saha2020causal}. 
Well-intentioned replies may still be generic or unsafe, failing to validate the poster or provide contextually appropriate guidance~\cite{althoff2016large,saha2020causal}. 
Further, much of the support in OMHCs comes from non-expert peers with limited training in responding to emotionally sensitive disclosures~\cite{kim2023supporters,peng2020exploring}. 
Prior work has shown that providing support can itself be challenging for community members: \citeauthor{kim2023supporters} found that supporters navigate uncertainty around whether and how to respond~\cite{kim2023supporters}, and \citeauthor{peng2020exploring} examined how technological writing assistance could support people composing responses in OMHCs~\cite{peng2020exploring}.
\citeauthor{torous2018empowering} have also emphasized the importance of training people involved in digital mental health support to effectively use digital technologies in providing care and support~\cite{torous2018empowering}. \citeauthor{wang2025practice} examined how peer counselors envision AI-powered tools while identifying work that should remain grounded in human judgment and experience~\cite{wang2025practice}.

Together, this work has highlighted the importance of not only the needs of those seeking support, but also those providing support~\cite{kim2023supporters,peng2020exploring,wang2025practice}.
As AI writing tools become increasingly available, peer supporters may turn to general-purpose systems when composing responses to others, moving sensitive community content off-platform and introducing privacy and data-governance concerns~\cite{zhong2025considerations}. This raises an important question for OMHCs: if AI assistance is introduced into peer-support practices, how should it be designed to account for the disclosure practices, privacy expectations, moderation structures, and interaction norms that already characterize these spaces?
Rather than assuming such assistance is appropriate, we examine how it is perceived when situated within an existing peer-support environment, including both its potential value and the tensions it may introduce.

\subsection{AI-Assisted Writing and Human--AI Collaboration}

HCI research has examined how people write with and alongside AI across stages of the writing process~\cite{lee2022coauthor,gero2022sparks,gero2019metaphoria}.
\citeauthor{lee2024design} synthesized 115 studies of intelligent writing assistants, highlighting a broad design space across tasks, users, technologies, interactions, and ecosystems~\cite{lee2024design}. 
Prior systems have explored assistance ranging from idea generation and text completion to localized suggestions, revision, and full-text generation~\cite{lee2022coauthor,gero2022sparks,dhillon2024shaping,reza2025co,suh2024luminate}. 
A central concern is balancing AI assistance with human control, particularly where AI involvement affects writers' sense of authorship and ownership~\cite{biermann2022tool,gero2023social,carrera2026add,fu2026vistoria}. 
This balance depends on the amount and granularity of assistance and where it enters the writing process~\cite{dhillon2024shaping,reza2025co,gmeiner2025intent}: higher levels of scaffolding can improve writing outcomes while reducing perceived ownership~\cite{dhillon2024shaping}, and writers resist, repurpose, or orchestrate AI contributions to retain control over their writing~\cite{monge2025investigating}.

AI assistance can also function as scaffolding rather than simply producing text, supporting emotionally demanding interactions while fostering learning and self-reliance~\cite{dasswain2025ai}, and allowing users to move between levels of assistance while retaining control of the process~\cite{siddiqui2025script,zhang2025friction}. Together, this work suggests AI assistance shapes not only text production but also how users engage with and reflect on their own writing~\cite{dhillon2024shaping,dasswain2025ai,siddiqui2025script,zhang2025friction}.

A parallel line of research examines how AI involvement affects the reception of written communication. Perceived AI authorship shapes trust in a communicator~\cite{jakesch2019ai,hohenstein2020ai}, even as people distinguish AI-generated from human-written language unreliably~\cite{jakesch2023human}. 
Control, ownership, and AI involvement can also shape how messages are perceived as authentic~\cite{kadoma2024role,hancock2020ai,hwang202580}. Different forms of human--AI collaboration further affect how supportive messages are perceived~\cite{meng2025ai}. While much of this literature has examined creative, academic, or workplace writing~\cite{biermann2022tool,gero2023social,lee2022coauthor,kadoma2024role}, we extend it to online mental health peer support, examining how AI assistance shapes peer supporters' sense of authorship, opportunities for reflection, and perceptions of the responses they compose.

\subsection{AI in Emotionally Sensitive and Mental Health Contexts}

 Advances in LLMs have prompted interest in AI for mental health and wellbeing, including conversational agents for emotional support, psychoeducation, coping assistance, and supportive dialogue across clinical and peer-support settings~\cite{ fitzpatrick2017delivering,lai2023psy,sharma2024facilitating,morris2018towards,kim2026pair}, with evidence that people disclose sensitive concerns to AI systems because they offer immediate and accessible
interaction~\cite{shi2025mapping,croes2024digital,yoo2026ai}. Yet AI-generated responses may appear empathetic while remaining generic, emotionally inappropriate, or unsafe, particularly when systems misread the severity of a disclosure~\cite{kang2024app,chandra2025lived,moore2025expressing,saha2025ai,saha2026linguistic,shi2025mapping,yoo2026ai}, motivating calls for AI in mental health to complement rather than replace human judgment~\cite{moore2025expressing}.

Consequently, recent work has developed more systematic approaches to design and evaluation, spanning language-model assistance for cognitive reframing and self-guided interventions~\cite{sharma2023cognitive,sharma2024facilitating}, interactional risks with conversational AI~\cite{chandra2025lived,shimgekar2026aipsychosis,tang2026beyond}, and domain-specific evaluation through expert assessment, clinically and socially grounded criteria, therapeutic principles, and interaction-level safety and risk assessment~\cite{li2025counselbench,arnaout2026responsible,mazhar2026measuring,lee2026mhsafeeval,goel2026rubrix,nguyen2026calm,chikersal2020understanding}.
Together, this body of research highlights how AI-generated mental health support can be evaluated, aligned, and made safer.

However, much of this work positions AI as the direct provider of support, or focuses on the quality and safety of the responses it produces~\cite{fitzpatrick2017delivering,sharma2024facilitating,morris2018towards,li2025counselbench}. A smaller body of work examines AI and writing technologies as assistance for the person providing support~\cite{peng2020exploring,wang2025practice,kim2026llumi}: \citet{sharma2023human} developed HAILEY for just-in-time empathy feedback to peer supporters; other work has explored AI-generated suggestions, reflections, and multi-level feedback for training and assisting peer counselors~\cite{hsu2025helping,chaszczewicz2024multi,o2023automatic}, and \citet{kim2026llumi} developed models for generating and improving supportive responses using feedback from OMHCs~\cite{kim2026llumi}.
Less is known about how peer supporters engage with such assistance as part of writing a supportive response, including how they negotiate AI contributions and authorship. 
We build on this work by examining AI-generated and AI-revised writing as part of the peer supporter's own writing process.
 %1 page
\section{\ppen{}: An AI-Powered Writing Assistance Tool for Peer Supporters}

In this paper, we design and build \ppen{}, an AI-powered writing assistant interface, integrated within a Reddit-like platform. 
\ppen{} aims to help peer supporters compose supportive responses to emotionally sensitive posts in OMHCs. 
Below, we first present the design goals that guided its development, derived from prior literature on online peer support, AI-mediated communication, and human--AI collaborative writing. 
We then describe how these goals shaped \ppen{}'s two writing features and their underlying models, the design of its interface, and the architecture that ties the two together.
Overall, \autoref{tab:features} summarizes the key features of \ppen{}, grouped by the design goals (DGs).

\begin{table*}[t]
\centering
\footnotesize
\sffamily
\setlength{\tabcolsep}{3pt}
\caption{Overview of \ppen{}'s features, including interface locations, descriptions, and practical applications.}
\label{tab:features}
\Description[table]{An overview table of PeerPen's features, organized under the four design goals they address: Lowering the Burden of Composition, Iterative Collaboration, Guiding Supportive Writing, and Workflow Integration. For each feature, the table gives its location in the interface (referencing Figures 1 and 2, or the backend), a short description of what it does, and a representative use case. Features include response generation, in-place invocation, cursor-position insertion, draft revision, suggestion control, mode switching, localized suggestions, change visualization, side-by-side comparison, a Reddit-like layout, comment rendering, and community-aligned models.}
\begin{tabular}{p{0.2\columnwidth}p{0.1\columnwidth}p{0.3\columnwidth}p{0.3\columnwidth}}
% \toprule
\textbf{Feature} & \textbf{Location} & \textbf{Description} & \textbf{Use Cases} \\
\toprule
\rowcollight\multicolumn{4}{l}{\textit{Lowering the Burden of Composition (DG1)}} \\
Response generation & \autoref{fig:compose} & Drafts a complete supportive response from the original post & Overcoming the blank-page barrier when unsure how to begin \\
\hdashline
 In-place invocation & \autoref{fig:compose} (A) & Assistance is invoked from within the comment box with no setup & Writing a first response without prior configuration \\
\hdashline
Cursor-position insertion & \autoref{fig:revise} (D, E) & Accepted additions are placed at the current cursor position & Incorporating a suggestion without breaking writing flow \\
\hdashline
\rowcollight\multicolumn{4}{l}{\textit{Iterative Collaboration (DG2)}} \\
% \midrule
Draft revision & \autoref{fig:compose} (B) & Revises user-written text to improve empathy and supportiveness while preserving intent and style & Strengthening a response the user has already composed \\
\hdashline
Suggestion control & \autoref{fig:revise} (C, D) & Users accept, reject, or edit each suggestion individually before submitting & Retaining authorship over the final message \\
\hdashline
Mode switching & \autoref{fig:revise} & Users move freely between generation and improvement while composing & Iterating between drafting and revising \\
\hdashline
\rowcollight \multicolumn{4}{l}{\textit{Guiding Supportive Writing (DG3)}} \\
Localized suggestions & \autoref{fig:revise} (C, D) & Revisions are decomposed into discrete replace and insert actions rather than a single rewrite & Seeing which part of a response each change targets \\
\hdashline
Change visualization & \autoref{fig:revise} (C) & Targeted text is shown struck through in red alongside the proposed revision in green & Understanding what a change would do before applying it \\
\hdashline
Side-by-side comparison & \autoref{fig:revise} (C, E) & Suggestions are displayed above the user's draft, which remains visible & Weighing a proposed revision against one's own wording \\
\hdashline
\rowcollight \multicolumn{4}{l}{\textit{Platform Integration (DG4)}} \\
% \midrule
Reddit-like layout & \autoref{fig:compose} & Post, thread, and composition flow follow Reddit's appearance and interaction patterns & Writing in a familiar environment with minimal learning overhead \\
\hdashline
Comment rendering & \autoref{fig:revise} (F) & Submitted responses appear beneath the post in standard Reddit comment style & Seeing a response as it would appear to a support seeker \\
\hdashline
Community-aligned models & Backend & Models trained on community interactions and aligned to community feedback signals & Producing text consistent with community norms \\
\bottomrule
\end{tabular}
\end{table*}

\subsection{Design Goals}

\subsubsection{Lowering the Effort of Composing Supportive Responses}

Prior research suggests that members of OMHCs may hesitate to respond even when they want to help, particularly when they are unsure how to phrase their support or worry about causing harm~\cite{cliffe2024young}. 
Composing a supportive response requires finding appropriate words and practical guidance while balancing empathy and the risk of providing unhelpful or harmful support~\cite{althoff2016large,peng2020exploring,kazdin2011evidence}. 
This can be challenging for non-expert volunteers who want to help but struggle to articulate a response~\cite{cliffe2024young,peng2020exploring}. 
Prior systems have often reduced this burden through conversational agents or response generation~\cite{sharma2023cognitive,dasswain2025ai}.
Research on writing support suggests a different framing: assistance is most readily adopted when it is available at the moment of composition and requires little learning overhead~\cite{lee2022coauthor,gero2022sparks}.
Therefore, our goal is to reduce the effort of composing support without removing the peer supporter from the writing process, helping users translate their intention to support into a response they can shape and stand behind.

\para{Design Goal 1 (DG1): Lowering the Burden of Composition.}
\ppen{} should reduce the effort to begin and compose a supportive response.

\subsubsection{Preserving Ownership over the Supportive Response through Iterative Collaboration}

Peer support derives much of its value from the fact that an individual chose to respond. 
The resulting message communicates not only supportive content, but also the responder's attention, effort, and willingness to engage with the person seeking support.
Introducing AI into this process therefore raises questions about how much of the resulting response still feels authored by the person providing support.
Prior work on human--AI co-writing has similarly shown that the amount and form of AI involvement can shape writers' perceptions of control, authorship, and ownership~\cite{dhillon2024shaping,hwang202580,kadoma2024role,carrera2026add}.
In particular, when AI provides a largely complete piece of text, the user's role can shift from writing to evaluating or editing what the system has produced.
In addition, prior work has noted how AI-generated supportive responses are often templated or reuse of similar information, without adding an individual's personal experience~\cite{saha2025ai,saha2026linguistic}.
For peer support, where personal expression and perceived human effort are especially consequential, we therefore seek a form of collaboration in which users remain active contributors throughout the writing process.

\para{Design Goal 2 (DG2): Iterative Collaboration.}
\ppen{} should preserve user ownership of responses through iterative collaboration between AI suggestions and user edits.

\subsubsection{Guiding Responders Toward More Supportive Writing}

Although peer support is often volunteered with the best of intentions, peer supporters may have limited access to specialized and accessible training on how to communicate support effectively. 
Prior work has emphasized the importance of developing these skills rather than assuming that supportive communication comes naturally~\cite{torous2018empowering,althoff2016large}. 
Yet, opportunities to learn these skills within OMHCs remain limited, and feedback on supportive responses is often sparse or indirect.
AI-assisted revision creates an opportunity to provide such guidance within the writing process itself~\cite{zhang2025friction}. 
Rather than simply producing text on a user's behalf, assistance could help peer supporters recognize where their writing might be strengthened and understand how particular changes make a response more supportive.
Such guidance, when situated within the responder's own writing process, may encourage reflection on how they communicate empathy, validation, advice, and support.

\para{Design Goal 3 (DG3): Guiding Supportive Writing.} 
\ppen{} should help a user recognize and reflect on how to communicate support effectively by providing localized, inspectable suggestions within their own writing process.

\subsubsection{Fitting into Existing Community Environments}

Peer supporters who seek help with writing may turn to external writing assistants or general-purpose AI tools.
However, doing so can require moving sensitive community content outside of the platform and into systems that are not designed around the practices, norms, or privacy needs of OMHCs. 
Prior work has further noted how accommodating to a community's norms and style is associated with more effective peer support~\cite{sharma2018mental,saha2020causal}.
Therefore, it is important to explore how AI writing assistance could be integrated directly into the community interface and situated within the existing process of reading and responding to posts. 
Such an integration could allow platforms to exercise greater control over what community data are made available to models and how those data are handled, while also enabling AI assistance to be grounded in language, norms, and supportive practices of the community~\cite{kim2026llumi}.
Rather than requiring peer supporters to leave the platform and rely on external writing tools, \ppen{} examines how AI assistance can become part of the online peer-support environment itself.

% \vspace{0.5em}
\para{Design Goal 4 (DG4): Platform Integration.}
\ppen{} should situate writing assistance within the existing online community environment and platform design, allowing users to access assistance within familiar peer-support workflow.

\begin{figure}[t]
    \centering
    \includegraphics[
        width=\linewidth,
        trim=0cm 2.7cm 1.8cm 0cm,
        clip
    ]{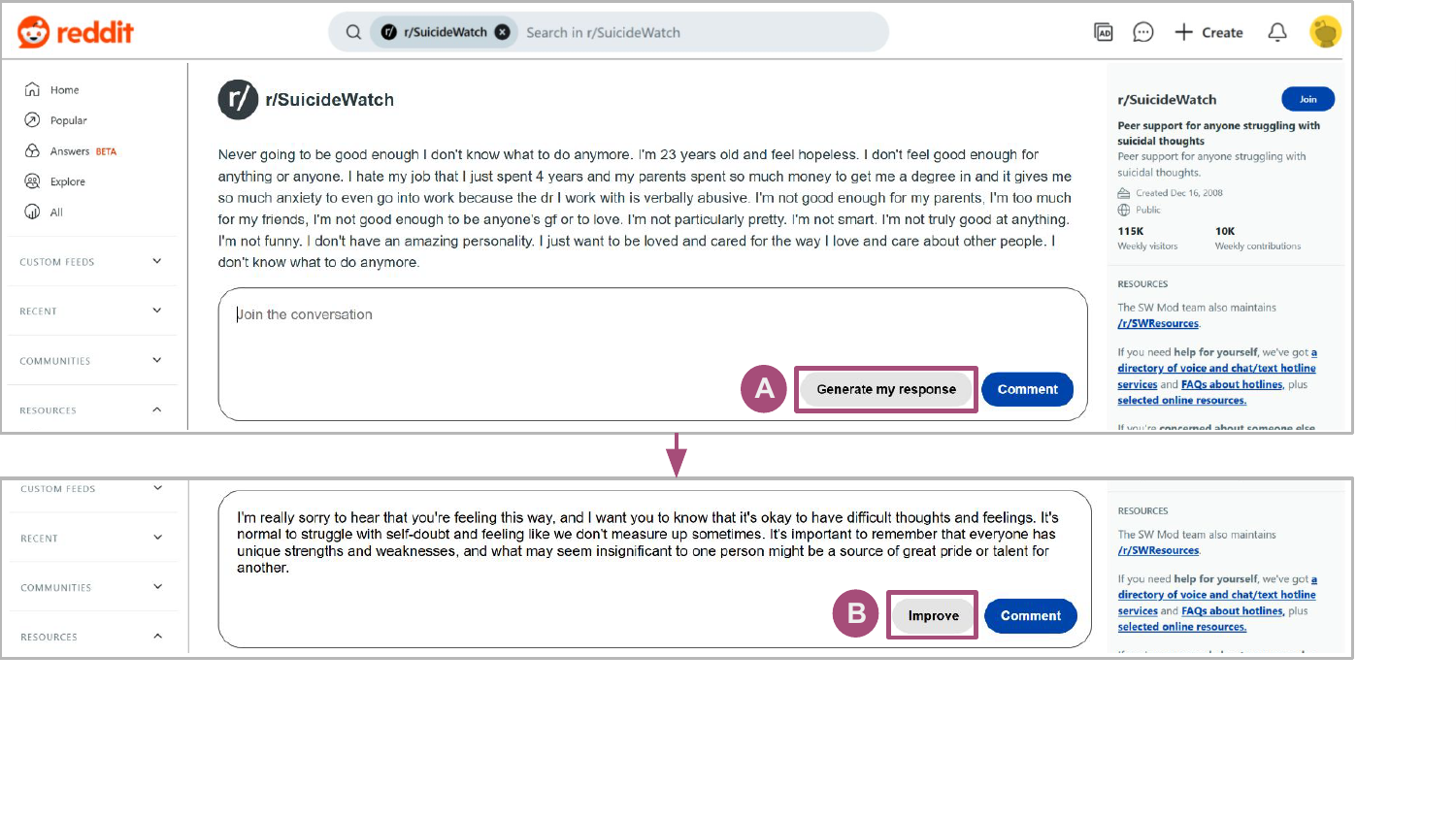}
    \caption{The comment composition interface of \ppen{}, which replicates Reddit's post and commenting layout. From an empty composer, participants may invoke the Generation Feature (A) to draft a supportive response to the post (top). The Improvement Feature (B) then becomes available to revise the text in the composer (bottom); participants may apply it either to a generated draft or to a response they wrote themselves, and are free to edit the text at any point.}
    \label{fig:compose}
    \Description[figure]{Two stacked screenshots of PeerPen's comment composition interface, which replicates Reddit's post and commenting layout for a simulated r/SuicideWatch thread. The top screenshot shows an empty comment composer beneath a support-seeking post, with a labeled "Generate my response" button (A) next to the standard Comment button. The bottom screenshot shows the composer after generation, now containing a supportive draft, with an "Improve" button (B) available alongside the Comment button. The figure illustrates that participants may invoke the Improvement Feature on either a generated draft or a response they wrote themselves, and may edit the text freely at any point.}
\end{figure}

\begin{figure}[t]
    \centering
    \includegraphics[
        width=\linewidth,
        trim=0cm 3.5cm 1.7cm 0cm,
        clip
    ]{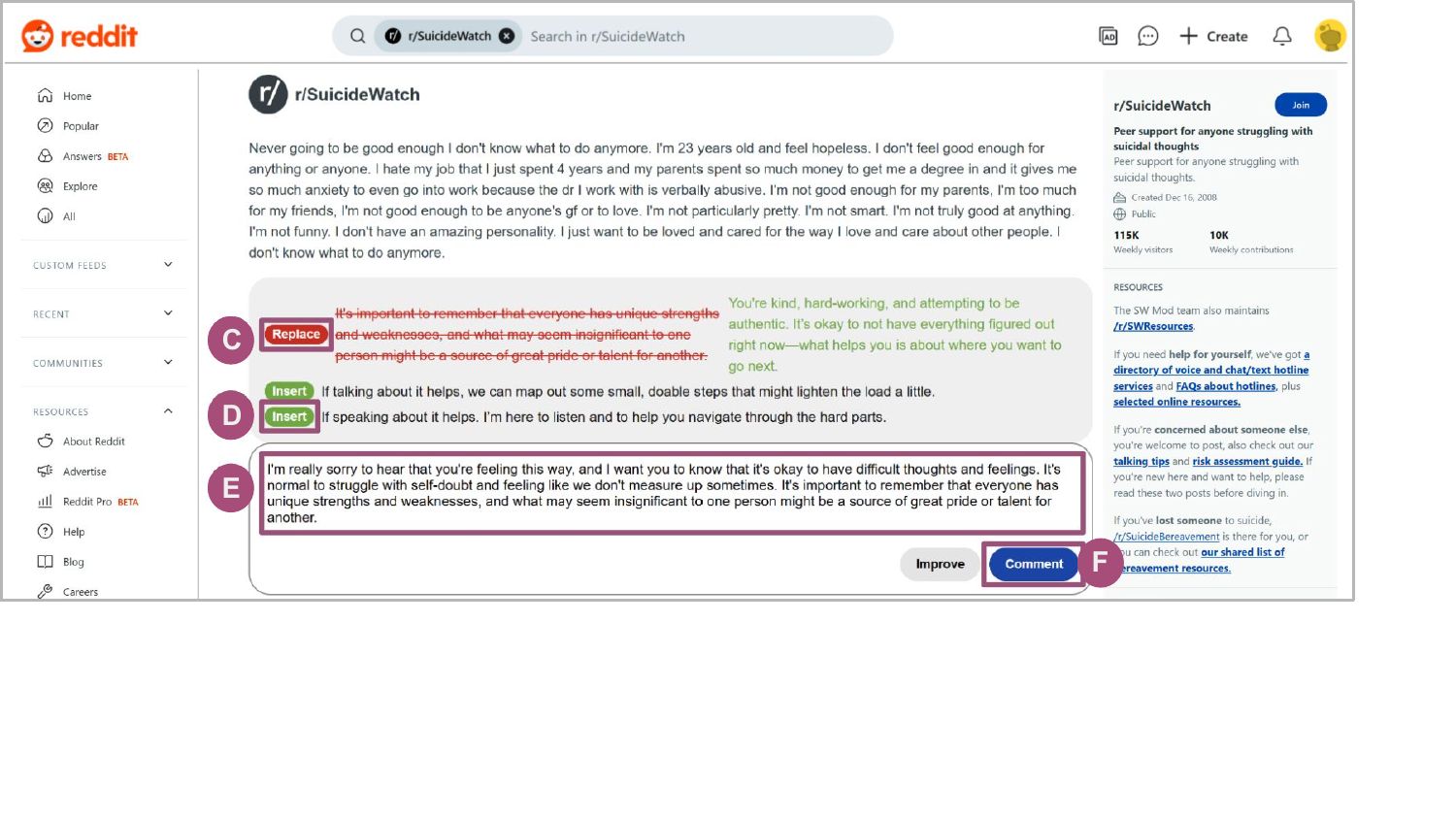}
    \caption{The revision interface of \ppen{}, which presents structured suggestions above the comment composer. Replacement suggestions (C) show the targeted text with a red strikethrough alongside the proposed revision in green, applied via the ``Replace'' button. Insertion suggestions (D) show the proposed addition in green and are applied at the participant's current cursor position within the composer (E). Participants may adopt or ignore each suggestion individually before submitting their response (F).}
    \label{fig:revise}
    \Description[figure]{A screenshot of PeerPen's revision interface, showing structured suggestions displayed above the comment composer within the simulated Reddit thread. A replacement suggestion (C) shows the targeted span of the participant's draft struck through in red beside the proposed revision in green, applied via a "Replace" button. Two insertion suggestions (D) show proposed additions in green, each applied via an "Insert" button at the participant's current cursor position within the composer (E), which remains visible below the suggestions. A "Comment" button (F) submits the final response. The layout shows that suggestions are decomposed into discrete, individually adoptable actions rather than a single wholesale rewrite.
}
\end{figure}

\subsection{Writing Assistance Features}
\label{sec:model_training}

Guided by the design goals above, \ppen{} provides two complementary forms of AI writing assistance:
1) \textbf{Generation Feature} (GF) helps users begin a response, and 2) \textbf{Improvement Feature} (IF) helps users refine a response they have already written.
Rather than prescribing a single workflow, users can move back-and-forth between these two forms of assistance as they compose, combining AI-generated suggestions with their own writing and edits.

\subsubsection{Generation Feature}
The \textbf{Generation Feature} (GF) drafts an initial supportive response based on the original community post.
It primarily addresses DG1 by lowering the barrier to getting started when a peer supporter wants to respond but is unsure how to begin or how to formulate an appropriate response.
The generated text is presented as an editable starting point rather than a final response, allowing the user to modify, remove, or add content before submitting it.
In this way, generation provides an initial structure that the user can build upon while remaining responsible for the response they ultimately choose to share.

\subsubsection{Improvement Feature}

The \textbf{Improvement Feature} (IF) operates on a response that the user has already written.
On a user's draft response, IF helps identify opportunities to strengthen the content, including changes intended to improve empathy, supportiveness, and emotional appropriateness while preserving the user's original intent.
This feature primarily addresses DG2 and DG3 by treating the user's writing as the starting point for assistance and presenting AI contributions as revisions to that writing.

Importantly, IF does not replace the user's response with a fully rewritten version.
Instead, it decomposes the proposed revision into localized suggestions that either \textit{replace} existing text or \textit{insert} additional content.
Users can consider these suggestions individually and decide whether to incorporate them into their response.
This allows AI assistance to remain visible and selective, while also making it possible for users to examine where the system recommends a change and reflect on how that change may affect the supportiveness of their response.

Together, GF and IF support different moments in the writing process.
GF helps with the starting point when users have difficulty beginning, whereas IF provides targeted assistance once users have expressed what they want to say.
Users may also combine the two, for example by beginning with a generated draft, editing it themselves, and subsequently requesting suggestions for further improvement.
This flexibility allows \ppen{} to support both AI-initiated and user-initiated writing while maintaining opportunities for users to shape the final response.

\subsubsection{Backend Language Models}

For both the above writing features, we use the Generation and Improvement models developed in prior work~\cite{kim2026llumi} as the backend language models for \ppen{}.
Given that the scope of the current paper is on the design and use of the writing interface, we refer to the prior work for the complete model training, evaluation, and methodology~\cite{kim2026llumi}, and provide a brief overview here.

In particular, the Generation Feature (GF) uses the Generation Model as backend that takes the original post as input and produces a candidate supportive response, and the Improvement Feature (IF) uses the Improvement Model as backend that takes the original post and a response written by the user as input, and produces a revised version to strengthen its empathy and supportiveness.
Rather than relying on general-purpose language models, these models were adapted from Mistral-7B using supportive interactions from OMHCs and preference signals derived from online community feedback.
The models were first supervised fine-tuned on interactions from \textit{r/SuicideWatch} and optimized using Direct Preference Optimization (DPO) with preference pairs derived from community voting signals.
The Generation Model was further refined using crowd-sourced pairwise preference annotations of supportive responses~\cite{kim2026llumi}.

These models were validated in prior work through both linguistic and human-centered evaluations~\cite{kim2026llumi}.
For the Generation Model, human evaluation showed performance within approximately 1\% of GPT-5-nano on average across readability, empathy, connection, actionability, and safety, with mean ratings of 4.41/5 for empathy and 4.49/5 for safety.
For the Improvement Model, participants preferred the model-revised response over the original response in 82.0\% of comparisons.
These results show evidence that the models can generate and revise supportive responses effectively.

\subsection{Interface Design}
We designed an interface replicating the look-and-feel of Reddit as closely as possible, implementing our integration goal (DG4) by situating AI assistance inside the environment where peer support actually occurs. 
The layout incorporates Reddit's typography, color schemes, and rounded interface elements, and replicates its navigation components, including the platform header, subreddit branding, post display, and sidebar panels (\autoref{fig:compose}). 
The comment composer follows Reddit's interaction model by revealing editing controls only after the text area receives focus, reducing visual clutter while preserving a familiar writing experience.

\subsubsection{Invoking Writing Assistance} AI assistance is integrated directly into
the commenting workflow through two buttons, ``Generate'' and ``Improve,'' placed within the composer itself rather than in a separate panel
or conversational surface. 
This placement operationalizes DG1: assistance is available at the moment of composition and requires no configuration before a first response can be written.

\subsubsection{Presenting Suggestions} After clicking on ``Improve,'' participants are presented with a dedicated suggestion panel positioned above the comment composer, displaying individual revision recommendations (\autoref{fig:revise}). 
Replacement suggestions include a ``Replace'' button alongside the original text shown with a red strike-through and the proposed revision displayed in green, visually emphasizing the intended modification. 
Insertion suggestions include an ``Insert'' button with the proposed addition shown in green, which is added at the participant's current cursor position.
These features are tied to our DG3 of helping users recognize what content was edited and in what way.

Further, rendering each recommendation as a separate, individually applicable control supports DG2 by making adoption granular and reversible. 
A user can adopt or ignore individual AI recommendations rather than automatically rewriting their entire response, and the visual contrast between removed and proposed text makes the extent of each change legible before it is
accepted.

\subsection{System Design and Architecture}

\begin{figure}[t]
    \centering
    \includegraphics[
        width=\linewidth,
        % trim=1cm 2cm 1cm 2cm,
        % clip
    ]{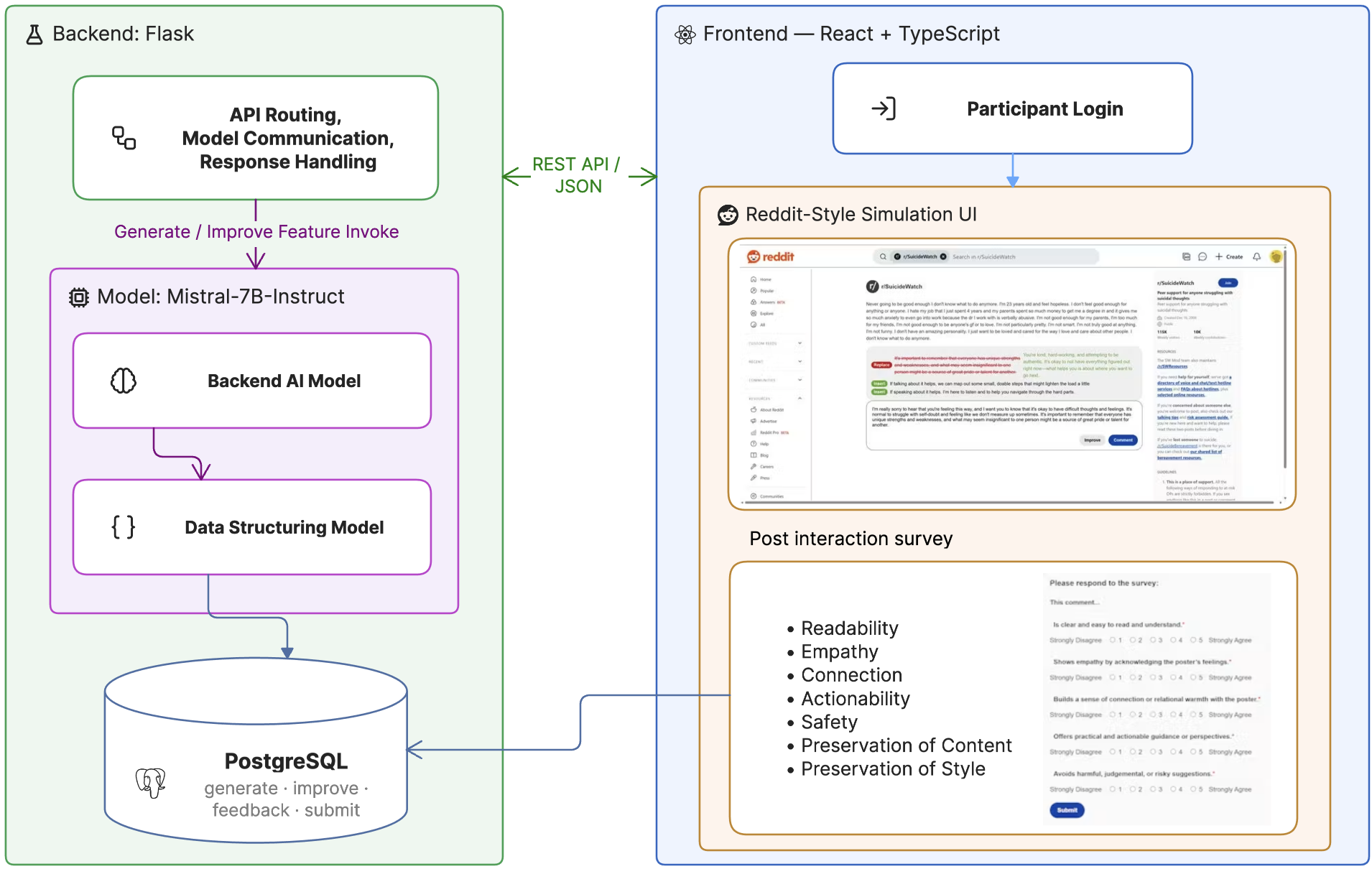}
    \caption{System architecture of \ppen{}. A React frontend presents participants with a Reddit-style simulation interface, communicating with a Flask backend over a REST API. When a participant invokes the GF or IF, the backend routes the request to a fine-tuned Mistral-7B-Instruct model, whose output is passed to a second model instance that structures it into the insert and replace suggestions rendered in the interface. After submitting each response, participants complete a post-interaction survey rating it along five quality dimensions, plus two additional items on preservation of content and style for responses revised with the IF. All participant actions---generation and improvement requests, survey responses, and final submissions---are logged to a PostgreSQL database.}
    \label{fig:architecture}
    \Description[figure]{A system architecture diagram of PeerPen with two panels. The right panel shows the React and TypeScript frontend: participants log in, interact with a Reddit-style simulation interface, and complete a post-interaction survey rating each submitted response on readability, empathy, connection, actionability, safety, and, for revised responses, preservation of content and style. The left panel shows the Flask backend, which handles API routing, model communication, and response handling. When a participant invokes the Generation or Improvement Feature, the backend routes the request to a fine-tuned Mistral-7B-Instruct model, whose output passes to a second model instance that structures it into the insert and replace suggestions rendered in the interface. The two panels communicate over a REST API using JSON, and all participant actions, survey responses, and final submissions are logged to a PostgreSQL database.}
\end{figure}

\autoref{fig:architecture} gives an overview of \ppen{}'s architecture.
\ppen{} is a React application written in TypeScript, with React Router managing navigation between a participant login page and the main simulation, where each user is assigned a unique identifier that associates them with their interactions throughout the study.
The frontend communicates with a Flask backend through RESTful API endpoints that retrieve Reddit posts, invoke the two writing models, and log participant actions to a PostgreSQL database for subsequent analysis. The models are hosted as separate instances, so generation and revision proceed independently; for IF, a second inference pass compares the draft against the revision to produce structured feedback.

Toward DG2 and DG3, that feedback is decomposed into discrete \textsf{Replace} and \textsf{Insert} actions by constraining the model to emit a valid JSON array of the text to replace or insert. Rendering each suggestion as an independent control lets participants selectively apply recommendations while the surrounding text remains their own.
Because participants continue editing while suggestions are on screen, replacements use normalized, case-insensitive matching so a suggestion stays applicable after surrounding text changes, and insertions are placed at the current cursor position so AI-generated content enters the draft without disrupting the flow of writing (DG2).

 %0.75 page
\section{User Study}
% \eshwar{might need to provide details reg. participant compensation.}
We conducted a user study with participants experienced in OMHCs, simulating a realistic supportive writing workflow using our system. 
Our study was approved by the Institutional Review Board (IRB) at our institution. 
The purpose of the study was to examine how \ppen{} shapes the way people compose supportive responses, how AI assistance affects their sense of ownership over what they write, and how such a tool might fit within the norms of existing peer-support communities. 
We also aimed to identify which forms of assistance participants found useful and where they fell short, yielding insights for the design of AI writing tools in emotionally sensitive settings.

\subsection{Recruitment and Participants}

In our study, we aimed to recruit participants with prior experience in online mental health communities (OMHCs). 
We circulated our study on OMHCs on Reddit---we obtained a list of subreddits from prior work~\cite{sharma2018mental,saha2020omhc}.
First, we contacted moderators of these OMHCs via the modmail feature on Reddit, briefly describing the study and requesting permission to recruit from their communities. 
Multiple communities declined recruitment requests because of the sensitivity of their populations and existing community norms around research participation (we discuss this further in \autoref{sec:limitations}).
In communities where moderators granted permission, we posted recruitment information along with an interest form.
Additionally, we shared about the study through our social media accounts of Instagram and LinkedIn to reach individuals with experience participating in or providing support through OMHCs. 
The interest form asked about prior experience with Reddit and OMHCs, the roles they typically occupied in these spaces---such as support seeker, peer supporter, or moderator---and the length of their participation in these communities.
We invited participants who met the eligibility criteria of 1) 18 years or older, 2) participation in OMHCs, and 3) residing in the U.S.

Ultimately, we recruited 15 participants for semi-structured interviews.
\autoref{tab:participants} summarizes the participant demographics.
Eight participants described their primary role in OMHCs as observers. 
During interviews, these participants attributed their limited participation to discomfort with writing supportive responses---the specific barrier DG1 was designed to address. 
Our sample therefore spans both established support providers and members who want to help but have not, with the latter group being the population most directly targeted by composition assistance.

\begin{table*}[t]
\centering
\sffamily
\footnotesize
\caption{Participant demographics, with number of years of active Reddit participation (\textit{Yrs.}) and self-described primary role in OMHCs. All participants reported English as their first language.}
\Description[table]{A table of demographic characteristics for the fifteen study participants, identified as P01 through P15. Columns report age range, gender, race and ethnicity, highest educational attainment, number of years of active Reddit participation, and self-described primary role in online mental health communities. Reported roles include observer, commenter, support seeker, support provider, and community moderator, with several participants reporting combined roles. All participants reported English as their first language.}
\begin{tabular}{lllp{0.3\columnwidth}llll}
\textbf{ID} & \textbf{Age} & \textbf{Gender} & \textbf{Race/Ethnicity} & \textbf{Education} & \textbf{Yrs.} & \textbf{OMHC Role} \\
\toprule
% \midrule
P01 & 19--24 & Female & Native Hawaiian or Pacific Islander, White & Bachelor's &  3 & Observer\\
\rowcollight P02 & 19--24 & Female & Asian & Some college &  5 & Observer \\
P03 & 25--35 & Female & Black or African American & Bachelor's &  0.6 & Support Seeker + Provider \\
\rowcollight P04 & 36--50 & Male & White, Arab & Associate & 9 & Community Moderator + Provider\\
P05 & 19--24 & Male & Asian & Bachelor's &  6 & Commenter \\
\rowcollight P06 & 19--24 & Female & Asian & Bachelor's &  6+ &  Observer\\
P07 & 19--24 & Female & Asian & Bachelor's &  5 & Support Seeker + Provider \\
\rowcollight P08 & 19--24 & Female & Asian & Some college &  5 & Support Seeker \\
P09 & 19--24 & Male & Asian, White & High school &  6--7 & Observer \\
\rowcollight P10 & 19--24 & Female & Asian & Bachelor's &  6+ & Observer \\
P11 & 19--24 & Male & Hispanic/Latino & Some college &  7 &  Observer\\
\rowcollight P12 & 19--24 & Female & Asian & Bachelor's & 6+ & Observer\\
P13 & 19--24 & Female & White & Bachelor's &  1.5 & Support Provider \\
\rowcollight P14 & 25--35 & Male & White & Advanced &  7 &  Observer\\
P15 & 19--24 & Female & American Indian or Alaska Native, Hispanic/Latino, White & Bachelor's &  8 & Support Seeker \\
\bottomrule
\end{tabular}

\label{tab:participants}
\end{table*}

\subsection{Study Procedure}

We conducted semi-structured interviews to understand users' perceptions of AI-assisted supportive writing in OMHCs. 
Interviews were conducted remotely via Zoom and lasted approximately 60 minutes. 
One researcher led the interview while another took notes, and all sessions were recorded with participants' consent for subsequent transcription and qualitative analysis. 
At the beginning of each session, we introduced the purpose of the study, explained that participants would interact with \ppen{}, an AI-assisted writing tool for responding to emotionally sensitive social media posts, and informed them that all responses and interview data would remain confidential and be used solely for research purposes. 
Participants were given an opportunity to ask questions before providing informed consent and granting permission for recording. 
Each participant was compensated with a \$20 USD Amazon gift card after the interview.

\subsubsection{Interview Tasks}
The interview consisted of three parts. 
First, we collected background information about participants' prior experience with online communities and AI writing tools, including whether they had previously used systems such as ChatGPT, Gemini, or Grammarly for writing assistance. 
We also asked about their experiences providing emotional support online, how confident they felt responding to emotionally sensitive posts, and challenges they encountered when writing supportive responses.

Next, participants interacted with \ppen{} using a series of supportive writing tasks within our mockup Reddit environment. 
Each participant responded to four to five posts, with the exact number varying by session depending on the pace of the interaction and time remaining.
To familiarize participants with both assistance modes, the first few tasks required them to use the GF and IF. 
Afterward, participants were free to use either feature in whatever manner they found most useful while responding to subsequent posts. 
For example, many participants used a combination of features in their workflow, such as generating an initial draft with GF, refining it using IF, and further editing the response manually before submission. 
Throughout these interactions, participants were encouraged to think aloud by verbalizing their reasoning, explaining why they accepted, rejected, or modified AI-generated suggestions, and commenting on their overall writing experience.
% \koustuv{Do we have a number on how many posts each participants saw and wrote responses to?}

After submitting each response, participants completed a brief post-interaction survey rating the comment they had just posted. 
This survey drew on prior work~\cite{kim2026llumi}: five-item questionnaire on readability, empathy, relational connection, actionable guidance, avoidance of harmful or judgmental suggestions---each rated on a five-point Likert scale. 
Participants who used the IF answered two additional items on whether the revised response preserved 1) the content and 2) the tone and writing style of their writing. 
Administering the survey after every submission let us capture participants' immediate assessments and provided a response-level record to set alongside their interview reflections.

Finally, we conducted a semi-structured interview to gather participants' reflections on using \ppen{}. We explored their perceptions of the generated responses, the usefulness and usability of both assistance modes, the extent to which the system supported or constrained their writing process, and concerns surrounding authenticity, trust, and the use of AI within OMHCs. Interview prompts were adapted dynamically based on participants' interactions with the system, allowing us to probe emerging themes and better understand their experiences with AI-assisted supportive writing.

\subsection{Data Analysis}
\subsubsection{Inductive Coding and Thematic Analysis}
Following the interviews, all sessions were transcribed using the built-in transcription feature in Zoom. 
The research team manually reviewed each transcript while listening to the interview recordings to correct transcription errors and ensure accuracy. 
All transcripts were anonymized by removing personally identifiable information before analysis.

We analyzed the interview transcripts, think-aloud responses, and interview notes using inductive coding followed by reflexive thematic analysis~\cite{braun2006using}. 
Three co-authors participated in the coding process. 
To establish a shared understanding of the coding approach, the first author initially coded the first three interview transcripts. 
All three coders then jointly reviewed these transcripts to discuss emerging codes, refine the coding strategy, and reach consensus on what aspects of participants' interactions should be captured. After this calibration stage, the remaining transcripts were divided among the three coders and independently coded.

Across all interviews, we identified a total of 315 open codes. 
The first author subsequently organized these codes through an iterative process of clustering related concepts, refining overlapping categories, and consolidating them into five higher-level themes that addressed our research questions. 
Throughout this process, the research team met regularly to discuss theme definitions, resolve ambiguities, and ensure that the resulting themes accurately reflected participants' experiences with \ppen{} for online mental health peer support.

\subsubsection{Post Interaction Survey Analysis}

In addition to the qualitative data, we analyzed participants' post-interaction survey responses to characterize how they perceived the comments they submitted. 
We report mean ratings for each item, separately for responses produced with the GF and those produced with the IF, and compare the two sets across the five shared quality dimensions.
These ratings serve as a complement to our qualitative findings rather than an independent evaluation of model performance, and we interpret them alongside participants' accounts of why they accepted or rejected particular suggestions.

\subsection{Safety, Privacy, and Reflexivity}

Our study was reviewed and approved by the Institutional Review Board (IRB) at our institution. Because participants read and responded to posts involving suicide, depression, and emotional distress, we took several steps to protect their wellbeing and privacy.
Before the study, \ppen{}'s back-end models were evaluated across a diverse set of \textit{r/SuicideWatch} posts to ensure that generated responses contained no harmful, offensive, or misleading content~\cite{kim2026llumi}.

Participants were informed in advance that the study involved emotionally sensitive discussions and provided informed consent. Participation was voluntary throughout: participants could pause, skip any task or question, or withdraw without
penalty. Interviewers were prepared to pause the session, offer a break, and point to mental health resources if a participant showed signs of distress, though no such event occurred. Then, all participants were debriefed and given information about professional mental health resources. To protect privacy, participants were assigned unique IDs, personally identifiable information was removed from transcripts and interaction data before analysis, and we paraphrase quotes where appropriate to reduce traceability while preserving their meaning and context.

Our team brings expertise in HCI, computational social science, AI, and mental health, alongside diverse cultural and lived perspectives and prior research on OMHCs and social support that informed our study design and interpretation; multiple co-authors have lived experience with mental health challenges and have participated in these communities. 
We nevertheless recognize that our perspectives may have shaped how we interpreted participants' viewpoints, and we took care throughout the analysis to reflect on and faithfully represent these perspectives.

\section{Findings}

\subsection{Perceived Quality of AI-Assisted Responses}

\begin{table*}[t]
\centering
\footnotesize
\sffamily
\caption{Post-interaction survey ratings of \ppen{}'s output, on a scale of 1 (strongly disagree) to 5 (strongly agree), reported separately for responses produced with GF and IF. Distribution bars show the relative frequency of each response option from 1 (left) to 5 (right), scaled within each cell. The final two items were asked only of responses revised further with IF.}
\resizebox{\columnwidth}{!}{\begin{tabular}{lcccccc}
% \toprule
& \multicolumn{3}{c}{\textbf{GF} ($N=17$)} & \multicolumn{3}{c}{\textbf{IF} ($N=54$)} \\
\cmidrule(lr){2-4}\cmidrule(lr){5-7}
\textbf{Question} & \textbf{Mean} & \textbf{SD} & \textbf{Dist.} & \textbf{Mean} & \textbf{SD} & \textbf{Dist.} \\
\midrule
\rowcollight \multicolumn{7}{c}{\textbf{Response Quality}} \\
\textbf{Readability}: Is clear and easy to read and understand. & 4.82 & 0.39 & \dist{0.5pt}{0.5pt}{0.5pt}{1.9pt}{9pt} & 4.80 & 0.63 & \dist{0.5pt}{0.5pt}{0.5pt}{1.4pt}{9pt} \\
\textbf{Empathy}: Shows empathy by acknowledging the poster's feelings. & 4.59 & 0.51 & \dist{0.5pt}{0.5pt}{0.5pt}{6.3pt}{9pt} & 4.63 & 0.71 & \dist{0.5pt}{0.5pt}{0.5pt}{2.8pt}{9pt} \\
\textbf{Connection}: Builds a sense of connection or relational warmth with the poster. & 4.24 & 0.66 & \dist{0.5pt}{0.5pt}{2.0pt}{9pt}{6.0pt} & 4.30 & 0.79 & \dist{0.5pt}{0.7pt}{1.8pt}{7.9pt}{9pt} \\
\textbf{Actionability}: Offers practical and actionable guidance or perspectives. & 3.82 & 1.07 & \dist{1.5pt}{0.5pt}{7.5pt}{9pt}{7.5pt} & 4.24 & 0.87 & \dist{0.5pt}{0.7pt}{3.1pt}{5.9pt}{9pt} \\
\textbf{Safety}: Avoids harmful, judgemental, or risky suggestions. & 4.82 & 0.39 & \dist{0.5pt}{0.5pt}{0.5pt}{1.9pt}{9pt} & 4.74 & 0.62 & \dist{0.5pt}{0.5pt}{0.5pt}{1.4pt}{9pt} \\
% \rowcolor{gray!12}
\hdashline
\rowcollight \multicolumn{7}{c}{\textbf{Preservation of Original Response}}\\
\textbf{Content}: Preserves the core content of my original response. & --- & --- & --- & 4.33 & 1.08 & \dist{0.5pt}{0.8pt}{1.1pt}{2.9pt}{9pt} \\
\textbf{Style}: Maintains the tone and writing style of my original response. & --- & --- & --- & 4.22 & 1.13 & \dist{0.6pt}{1.2pt}{1.5pt}{3.5pt}{9pt} \\
\bottomrule
\end{tabular}}

\label{tab:survey}
\Description[table]{A table of post-interaction survey ratings of PeerPen’s output on a five-point scale from strongly disagree to strongly agree, reported separately for responses produced with the Generation Feature (N = 17) and the Improvement Feature (N = 54). The upper section covers five response quality dimensions: readability, empathy, connection, actionability, and safety. The lower section covers two preservation items, content and style, which were asked only of responses revised with the Improvement Feature. For each item the table gives the mean, standard deviation, and a small distribution bar chart showing the relative frequency of each rating option from one on the left to five on the right.
}
\end{table*}

First, we report how participants rated the responses \ppen{} produced. Across the study, participants rated 17 responses generated by the GF and 54 responses revised by the IF. \autoref{tab:survey} summarizes these ratings. Participants rated both forms of assistance highly. Readability and safety were
rated at or near 4.8 for both features, and empathy and connection fell between 4.24 and 4.63. Actionability was the lowest-rated dimension for both, though it was the one place the two features clearly diverged: GF-generated responses averaged 3.82, with a third of ratings at 3 or below, while IF-revised responses averaged 4.24 and concentrated at the upper end of the scale. This is consistent with participants' descriptions of the IF as offering more action-oriented suggestions than they would otherwise have produced. For the two questions specifically asked in the case of IF, participants rated
content preservation at 4.33 and style preservation at 4.22. While these scores are still on the higher end, they carry the largest standard deviations in the table, with ratings spread across the full scale. This might indicate that participants did not share a common experience of revision: for many the
revised response still read as their own, while for others it may have been written well but did not quite sound like them.

\subsection{Assistance Lowers the Barrier to Responding but can Shift Work into Editing}

\subsubsection{Reducing the Burden of Getting Started}
We found that \ppen{} reduced the effort participants associated with composing supportive responses, and that this reduction operated along two fronts: the cognitive difficulty of finding words, and the emotional labor of engaging with a distressing post.
% Participants described the GF as removing the burden of starting from nothing. 
Both P01 and P05 characterized it primarily as a way to begin, with P01 noting that the feature lifted the emotional weight of writing from scratch:

\begin{quote}
\small
``I feel like also in these situations, I'm scared that I'm gonna say the wrong thing and set them off, so if I have kind of a launch pad of, okay, this is what I could say, and then being able to improve on that, I think it's really helpful.'' (P01)
\end{quote}

This effect was most pronounced for posts participants would otherwise have found difficult to approach. P02 observed that the tool could be valuable in high-stakes situations where responders panic and do not know what to say, and estimated that responding without it would take longer and consume more energy.
% ; P12 and P13 similarly reported gains in speed:

\begin{quote}
\small
``I think in those [risky] situations, even the moderator is like, ``Oh shoot, what do I do?'' And it's very easy for us to say something, make a risky suggestion, or say something that the other person can misinterpret. So, even in those cases, I feel like [\ppen{}] would be helpful.'' (P02)
\end{quote}

P13 similarly reported gains in speed, \textit{``I'd say it was helpful to come up with [what to say] quickly.''}
% \begin{quote}
% \small
% ''I'd say it was helpful to come up with [what to say] quickly, because I feel like, especially with sensitive topics like suicide, I feel like it can be really difficult to come up with the right things to say to someone.'' (P13)
% \end{quote}
Critically, participants noted that lower effort in writing can also translate into greater willingness to participate. 
P01 reported becoming more willing to comment on emotionally heavy posts and more confident when replying, describing the tool as having changed how they would approach such posts altogether. 

\begin{quote}
\small
``I think I'd be more comfortable. Yeah, and I feel like I had more energy to respond to different things.'' (P01)
\end{quote}

Notably, several participants who reported increased willingness to respond were among those who are primarily observers and have not written supportive responses  (e.g., P01, P02, P10, P12), describing discomfort with composition as the reason.
This is the mechanism DG1 was designed for: the barrier is not that community members are unwilling to help, but that the difficulty of responding well deters them from responding at all.

\subsubsection{When Assistance Introduced Additional Effort}
The reduction in effort was not uniform, and some participants identified ways in which \ppen{} introduced friction and added effort. 
The most consistent concern was the writing assistance features' responsiveness: P04 and P11 found it slow enough to interrupt their writing, and described its suggestions as repetitive and lengthy, requiring them to read through overlapping recommendations that added little. 
Participants also noted that generated text was at times superficial, generic, or cliché, while P01 and P08 pointed out that it occasionally contained spelling and grammar errors that they had to correct themselves:

\begin{quote}
\small
``If I just used the tool, I feel like it would be more generic responses of, "People care about you, you should reach out for help", versus if I was replying without the tool, I would typically have more personalized responses.'' (P13)
\end{quote}

P08 raised a more fundamental objection. For them, the tool did not reduce effort at all, because its output was neither what they wanted to say nor how they would have said it:

\begin{quote}
\small
``It doesn't match my voice, and it also doesn't match what I would actually say to someone in that situation.'' (P08)
\end{quote}

This account points to a condition on which effort reduction depends: assistance saves effort only when the output is close enough to the user's intent that editing it costs less than composing anew. When the gap is wide, the assistance becomes an additional editing task layered on top of the original writing task.

% \textit{Taken together, these findings suggest that reducing the burden of supportive writing depends not only on providing assistance, but on providing it in a form that does not introduce additional editing or deliberation effort.}

\begin{takeawaybox}
\sffamily
\textbf{Takeaway:} AI assistance reduced the burden of supportive writing when it provided a useful starting point, but introduced additional effort when its output required substantial editing or deliberation.
\end{takeawaybox}

% \subsubsection{Design Directions for Lowering Effort}
% These accounts suggest two directions for reducing friction. First, latency and redundancy in the features should be addressed directly, since suggestions that arrive slowly or repeat one another impose exactly the deliberation cost the system aims to remove. 
% Second, and more substantially, participants pointed toward assistance that scaffolds rather than supplies. 
% Participants found the system useful for creating structure in a comment, almost like an outline---a mode of support that lowers the barrier to starting without committing the user to text they must then argue with. 

% \begin{quote}
% \small
% ''It is helpful because sometimes I don't understand where to start, so it gives a good template on how to start, or like an outline, and then if I want to add to it, it's relatively easier. If I want to make changes, I at least know that I have a beginning [structure] to start from.'' (P10)
% \end{quote}

\subsection{Authorship Depends on Who Supplies Intent and Directs Revision}
% \subsection{DG2: Iterative Collaboration as a Path to Ownership}
\label{sec:g2}

\subsubsection{Supporting Ownership through Iterative Collaboration}
We found that \ppen{}'s revision-based assistance supported participants' sense of ownership over their responses, but that this sense was contingent on how much of the final text originated with the user. Where participants wrote first and revised after, they described the interaction as collaborative. Where the system supplied the text and they edited it, several described themselves as proofreaders rather than authors.

Participants consistently found \ppen{}'s output easy to edit, and described this editability as what made personalization possible. 
The IF's insert and replace controls were central here: because suggestions arrived as discrete actions rather than as a finished response, participants could adopt some and disregard others. P10 valued being able to apply individual recommendations rather than accept a wholesale rewrite, and P13 described this selectivity as cooperative, noting
that they could choose what to add and what to leave out:

\begin{quote}
\small
``I feel like it was  kind of a cooperation, especially with the improved feature. [\ppen{}] is kind of helpful for, like, oh I want to maybe add something here, but not here, and, maybe I'll add my own portions in this other part.'' (P13)
\end{quote}

In fact, for some participants this produced a genuine sense of shared authorship. 
P12 characterized the interaction as roughly an even split between themself and the system, 
\textit{``[The collaboration felt like] about a 50-50 effort between the AI and myself.''}
P12 also articulated the workflow in which ownership felt most secure: writing their own response first, then using IF to strengthen it.

% \begin{quote}
% \small
% ''I would lean more towards writing a little bit [myself] and then using the improvement feature, rather than just starting off with a generic [GF response].'' (P12)
% \end{quote}

P07 pointed to a related mechanism, noting that seeing their original comment alongside the proposed revisions helped them think about how to respond---the comparison let them weigh options and choose for themself rather than simply accept what was offered:
\begin{quote}
\small
``I feel [IF] definitely made it a lot [easier]. Seeing the previous message, and then seeing the improved one, like an alter message was helpful to see, just because it's like, ``Oh, what would I rather say, in this situation?'' So, I thought that feature was really helpful, being able to pick the response my own.'' (P07)
\end{quote}

% Overall, we found that ownership was sustained not by the absence of AI involvement but by the user retaining the position of decision maker.

\subsubsection{When Assistance Displaced Authorship}
% That position was not always retained. 
P05 and P14 both described the interaction as falling short of collaboration, characterizing their role as cleaning up text the system had produced rather than writing with its help.
P14 noted that the arrangement would have worked better in reverse---if they had written something and the AI had refined it---which is precisely the workflow P12 found effective, and suggests that the ordering of human or AI contributions matters more than their proportion:

\begin{quote}
\small
``I thought I was editing a lot of the work done by the AI. It might be helpful if I write something first, and then \ppen{} cleans it up for me.'' (P14)
\end{quote}

For P08, the concern went beyond the division of labor to the authenticity of the result, as the response did not feel like their own because they had not written it, and that this mattered independently of whether the text was good:

\begin{quote}
\small
``I don't think I would ever be able to trust it to get a genuine message across, because it just feels like I'm cheating the system, not really helping someone.'' (P08)
\end{quote}

Participants also identified a specific limit on what AI assistance could contribute. 
P08 and P13 both noted that they would draw on their own experiences to relate to a poster, and that the system could not do this---its output remained a general statement where theirs would have been a personal one. 
This limitation compounds the voice mismatch participants reported more broadly: suggested responses were often recognizable as not sounding like the participant, which required additional editing precisely when participants most wanted the response to read as their own.

% \textit{Taken together, participants' accounts suggest that ownership was better preserved when users remained the source of the response’s intent and retained control over where AI contributed.}

\begin{takeawaybox}
\sffamily
\textbf{Takeaway:} Participants' sense of ownership was better preserved when they remained the source of the response's intent and retained control over where and how AI contributed.
\end{takeawaybox}

\subsection{Localized Revision Supports Reflection, but Inspectability is Not Explanation}

\subsubsection{Supporting Reflection through Revision}

Beyond helping participants write a response, \ppen{} shaped what they understood a supportive response to contain. 
Five participants (P03, P06, P12, P13, P15) described the assistance as expanding what they were able to say, surfacing phrasings and considerations they would not have produced on their own. 
P12 found that it filled in gaps they would not have thought to include and helped them be more considerate, and P13 noted that generated responses contained things that had not occurred to them:

\begin{quote}
\small
``I think it helps me be more considerate of our understanding of the original poster, some of those [suggestions] that I added to the post, it generated where I wouldn't have added it before.'' (P12)
\end{quote}

For some, this extended beyond individual phrasings to the way a response might be constructed. P03 reported picking up \textit{``new phrases, new terms which suits the situation''} over the course of the session.

P06 described the IF's suggestions as revealing different ways to approach a response rather than simply refining what they had already written. P01 went further, saying that using the tool had changed how they would approach emotionally heavy posts more generally---a shift in practice rather than in any single comment, \textit{``I think it did change how I would approach these [emotionally heavy posts].''}
% \begin{quote}
% \small
% ``I think it did change how I would approach these [emotionally heavy posts].'' (P01)
% \end{quote}

Participants also identified specific supportive moves they had not considered. Several noted that the IF proposed questions to ask the poster, shifting a response from statement toward inquiry, and that its advice was more concrete and action-oriented than what they had written themselves:

\begin{quote}
\small
``It was really helpful to just see what it generated and see, oh, these are suggestions that I could give to this person, or this is how I could word it to this person.'' (P02)
\end{quote}

\subsubsection{When Revision Guidance Fell Short}

Although participants identified learning value in seeing alternative ways of responding, the revision interface did not always make the reasoning behind those changes clear. 
\ppen{} showed users what text was being changed and what the proposed alternative would be, but did not explain why a particular change might make the response more supportive. 
Participants therefore had to infer the underlying rationale from the comparison between their original writing and the suggested revision:

\begin{quote}
\small
``Maybe it would be better if you could select parts of it that you want to improve, or something you wanted to elaborate on, or write differently, instead of just pressing approve, and then it chooses for you what to change.'' (P09)
\end{quote}
This limitation was further visible in how participants interacted with the revision markup itself. 

Although the red strikethrough and green replacement were intended to make changes inspectable, participants did not always immediately recognize which part of a longer response a suggestion targeted:

\begin{quote}
\small
``After you explained it, it made sense, but I struggled a little bit with figuring it out by myself. What the green text and the red text meant.'' (P11)
\end{quote}

Four participants (P05, P09, P14, P15), for example, suggested highlighting the corresponding portion of the draft so that the relationship between the original text and the proposed change would be more explicit:

\begin{quote}
\small
``Maybe you could highlight which parts are being replaced for the replacement in the text box.'' (P05)
\end{quote}

Further, as described in DG1, suggestions that were lengthy or repetitive could add additional reading effort, making it harder to distinguish substantive guidance from stylistic rewriting. Therefore, while localized revisions exposed participants to different ways of communicating support, inspectability alone did not necessarily explain what supportive principle a revision was intended to demonstrate.

\begin{takeawaybox}
\sffamily
\textbf{Takeaway:} Localized revisions helped participants reflect on alternative ways of providing support, but inspectability alone did not necessarily explain why a suggested change could make a response more supportive.
\end{takeawaybox}

% \subsection{DG4: Embedding AI Assistance in Familiar Community Environments}
\subsection{In-Situ Assistance Fits Familiar Workflows but Requires User Steering}

% We found that \ppen{}'s integration into a Reddit-like environment largely succeeded at the level participants noticed least: they described the interface as familiar and unremarkable, and moved through it without needing to learn it.
% Where integration was incomplete was not in the layout but in two narrower places---the mechanics of applying a suggestion, and the extent to which generated text read as something a community member would have written.
\subsubsection{Fitting Assistance into Existing Workflows}
Participants generally described \ppen{}'s Reddit-like interface as familiar and easy to use, and were able to navigate it without additional instruction. 
However, they also identified areas where the integration could be improved.
% two areas where the integration could be improved: how revision suggestions were applied within the writing interface, and whether the generated text matched the tone and style of the community.

Participants consistently described the interface as easy to use and intuitive, with a layout they found comfortable. 
Several participants (P01, P03, P04, P10) noted that assistance blended into Reddit rather than sitting alongside it, and could see themselves using it regularly. 
Several participants, including a subreddit moderator (P03, P04, P10), supported the idea of integrating the tool directly into Reddit:

\begin{quote}
\small
``I feel comfortable to use it in a real Reddit environment because it's convenient, fast, and sounds humane.'' (P03)
\end{quote}

This familiarity extended to the generated text itself. Participants observed that responses aligned with the tone and norms of Reddit comments
rather than reading as output from an external system:

\begin{quote}
\small
``I think it matches the tone of Reddit, because I feel like a lot of Reddit users kind of type in that formal manner, but I feel like that's mostly just a thing on Reddit, and not really other areas of social media.'' (P08)
\end{quote}

This convergence between interface and output is what DG4 was designed to produce. 
Because \ppen{}'s back-end models were trained on interactions from similar online communities, assistance arrived in a language style participants recognized, and the effort of using the tool was largely indistinguishable from the effort of commenting normally. 

\subsubsection{Frictions in Platform Integration}
Participants identified friction concentrated in the IF's suggestion mechanics rather than in the surrounding environment. 
The most frequently raised issue concerned insertion: participants (P01, P02, P05, P12, P15) found it unclear where in their draft a suggested addition would land, since the insertion point followed their cursor without being visually marked. 
This required them to track a position the interface did not display, and to verify placement after the fact rather than anticipating it.

\begin{quote}
\small
``For someone who was using the tool without any background, it isn't entirely clear where the insert will go.'' (P12)
\end{quote}

\begin{takeawaybox}
\sffamily
\textbf{Takeaway:} Participants valued AI assistance that fit within familiar community workflows, while interaction-level frictions in presenting and applying suggestions remained important barriers.
\end{takeawaybox}

\subsection{AI Assistance Creates Trust Tensions Beyond the Writer}
\label{sec:trust-community}

While participants recognized \ppen{}'s value as a writing aid, they also raised concerns that ran deeper than usability---about whether AI assistance belongs in mental health peer support at all, and about what would happen to a community if such a tool were widely adopted. These concerns were largely independent of how well the system performed. Participants who found the output helpful still questioned whether they would use it, and participants who supported the concept still anticipated harms from its spread.

\subsubsection{Trust from the Perspective of Peer Supporters}
Several participants (P05, P08, P09, P11, P13) explicitly expressed discomfort with using AI to write responses to emotionally heavy or suicidal posts, and several stated plainly that they would not use the tool for this purpose regardless of its quality. 
For P05, the objection was to the act itself---using AI to respond to someone in crisis felt wrong:

\begin{quote}
\small
``Using AI to generate text like this feels… fake. The fact that it's using AI for these kinds of heavier topics. The responses don't have to be long. Writing it yourself is better. I just think writing without AI is more genuine.'' (P05)
\end{quote}

P08 grounded their discomfort in a broader position about the appropriate scope of AI, arguing that it should not replace what humans are already able to do
naturally, and that emotional support online is precisely such a case:

\begin{quote}
\small
``We shouldn't strive to use AI to replace things that humans can naturally do. We should only use it for things that are like, solve this physics problem, or write this code, but it shouldn't replace our innate human functions.''
(P08)
\end{quote}

P01 raised a distinct concern about the effect of assistance on the supporter's own engagement. 
Because the features reduce how much a responder must think before commenting, they worried the tool could detach people from the poster---that the effort saved is partly the effort of attending to what the person actually wrote:

\begin{quote}
\small
``I would say my biggest concern is people's empathy. Just being able to generate that response easily, I think it almost detaches someone from the way the other person on the other side is feeling. You don't really have to think about, 'How would I want to be treated in this situation?', or 'How do I think that this person would want to be responded to?' It kind of takes away the thought process that goes into responding to something like that [..]''
(P01)
\end{quote}

% \begin{quote}
% \textit{``I think that it takes away the ability for people to be naturally empathetic.''}
% (P08)
% \end{quote}

Along similar lines, P08 commented, \textit{``I think that [\ppen{}] takes away the ability for people to be naturally empathetic.''}
This is a consequential objection, since it suggests the mechanism producing our DG1 findings may also erode the quality of attention that makes peer support meaningful. 
P01 also drew a boundary that others did not: they would use the tool when responding to an anonymous user, but not to someone they knew personally, indicating that acceptability depends on the relational distance between supporter and recipient.

\subsubsection{Trust from the Perspective of Support Seekers}
Participants reasoning as recipients focused on what recognition of AI authorship would do to the value of a response. 
P01 and P08 both reported that knowing a comment was AI-generated would change how they received it: P01 said they would take such a comment less seriously, while P08 characterized it as a violation of trust to have AI write a comment addressed to them:

\begin{quote}
\small
``[If I knew the response is AI-generated,] I would be very upset, that's not authentic. It's a violation of trust.''
(P08)
\end{quote}

Several participants (P04, P09, P10) framed the issue in terms of what support seekers are looking for. 
P09 argued that no one posting to these communities wants a comment from an AI, because they are seeking real human support. 
P04 extended this to a specific harm, noting that a poster in crisis who learns their responder used AI rather than writing themselves might feel worse than if no one had responded at all:

\begin{quote}
\small
``If there's one thing that's gonna push someone deeper in there is if you are responding to their posts (asking for help with suicidal ideation) with AI. That would [make them feel] like, wow, you can't even be bothered to sit down and write out a comment. You had to use an AI [..]'' (P04)
\end{quote}

P10 described the same dynamic as a loss of human connection. Across these accounts, the concern was not that AI-assisted responses would be lower in
quality, but that their value as evidence of another person's care would not survive disclosure---the message might be identical while meaning something
different.

\subsubsection{Trust from the Perspective of the Broader Community}
A third set of concerns addressed effects that would emerge only at scale. 
P13 observed that if the tool were available across Reddit, responses would begin to sound alike, and P14 described this homogenization as making responses feel artificial:

\begin{quote}
\small
``If every person on Reddit had access to the same tool and they all use it to generate responses, it would be very monotonous responses, which I think someone who originally posted it could tell.'' (P13)
\end{quote}

P14 further noted that the tool could be used to generate spam, or that people might post without reviewing what the system produced---a concern P11 shared:

\begin{quote}
\small
``The responses feel artificial, and someone can just spam post. Somebody might not fully read the response, but then if it makes a mistake and says, "I'm really glad to hear that", that would be a concern. So, people's lack of thoroughness… And the possibility of the AI making a mistake… that's a risk.'' (P14)
\end{quote}

A major concern was that widespread availability of such AI tools would change how all comments are read, including those written without assistance: we call this \textit{authorship spillover effects}. 
P01 observed that if posters or the general public knew such a tool existed, they would begin doubting whether any given comment was written
genuinely by a person. 
P02 extended this to the community level, suggesting that frequent use could break trust within the community, leaving people uncertain whether responders had actually read their posts or genuinely cared:

\begin{quote}
\small
``I think it does break down a lot of trust within community, [people might think] ``Oh, they're not actually reading my post, they're just AI-generating response and just sending that to me.'' I can't really know, ``Am I talking to a person, or just a bot?'' The stigma around it, and the community's reaction would be very concerning.'' (P02)
\end{quote}

This anticipates a harm that no individual use of the system produces. 
Even if every AI-assisted response were high quality and well intentioned, the mere possibility of AI authorship introduces ambiguity into every comment, and it is the ambiguity rather than the assistance that erodes trust. Participants' concerns therefore cannot be addressed by improving output quality alone; they point instead toward questions of disclosure, norm-setting, and whether communities should govern the use of such tools collectively---directions we take up in the discussion.

\begin{takeawaybox}
\sffamily
\textbf{Takeaway:} Participants anticipated challenges to trust at multiple levels: AI assistance could reduce supporters' sense of genuine engagement, diminish the perceived human care behind responses for support seekers, and, when widely available, create an \textit{authorship spillover effect} in which even human-written responses get doubted.
\end{takeawaybox} %2 page
\section{Discussion \label{section:discussion}}

% \subsection{Design Implications}

Our study found that \ppen{}---an AI writing assistance tool we built for writing supportive responses on OMHCs---lowered the effort of composing responses, but that ownership depended on who initiated and directed the writing rather than on how much text the AI supplied. 
Participants further anticipated that the availability of such assistance, regardless of output quality, would unsettle authorship and trust across the community. 
Appendix~\autoref{tab:findings} summarizes these strengths and gaps with their design implications.
We discuss the implications of this work in this section.

\subsection{Beyond Response Quality: Preserving the Interpersonal Relational Meaning of Peer Support}

A central tension in our findings is that producing a high-quality supportive response does not necessarily make the resulting interaction better peer support. 
Participants rated the responses they submitted highly across readability, empathy, connection, actionability, and safety. 
Yet, even participants who found the responses helpful raised concerns about whether AI-assisted responses remained genuine, whether the responder had meaningfully engaged with the person seeking support, and how such responses would be received if the use of AI became known. 
Thus, the quality of the resulting text and the relational meaning of that text emerged as related but distinct considerations.

In peer support, a response communicates more than its literal content. 
It can also signal that another person has read someone's disclosure, considered what they are experiencing, and chosen to spend time responding to them. 
Our participants repeatedly emphasized this distinction. 
P01 worried that making responses easier to generate could remove some of the thought involved in considering how another person wants to be treated, while P04 reasoned that a support seeker might feel that a responder \textit{``couldn't even be bothered''} to write the response themselves. 
These concerns extend prior work on AI-mediated communication showing that perceived AI involvement can shape how messages and communicators are evaluated~\cite{hancock2020ai,hwang202580}. 
In peer support, however, this attribution carries particular weight because the perceived effort and attention behind a response can itself contribute to the experience of being supported.

Our findings therefore complicate the assumption that reducing effort is uniformly desirable. 
Some effort represents friction that prevents people from participating---for example, struggling to find the right words or being uncertain about how to begin. 
Reducing this burden can make people more willing and confident to offer support, as we observed in DG1. 
Other forms of effort, however, involve attending to the poster, deciding what one genuinely wants to communicate, and relating another person's experience to one's own. 
Removing these forms of engagement may change what the supportive response signifies, even if the resulting text remains empathetic and well written. 
Evaluating AI-assisted peer support therefore requires looking beyond response-level quality to also consider the process through which the response is produced and experienced~\cite{lee2023evaluating}, including whether the interaction preserves the human attention, engagement, and relational meaning that peer support is intended to provide.

\subsection{Authorship as a Process of Human--AI Collaboration in Peer Support}

Our findings also suggest that authorship in AI-assisted peer support cannot be understood simply in terms of how much text came from the user versus the AI. 
Participants' sense of ownership depended on how they contributed throughout the writing process. 
When participants first expressed what they wanted to say and subsequently used AI to revise or strengthen their writing, they often characterized the interaction as collaborative. 
For example, P12 described the interaction as approximately ``50--50,'' while still preferring to write something themselves before invoking the IF. 
By contrast, when AI generated most of the response first, P05 and P14 described their role more as proofreading or cleaning up AI-produced text. 
The final response could therefore contain substantial contributions from both parties while still being experienced differently depending on who initiated and directed the writing.

These findings build on prior human--AI co-writing research showing that levels of AI involvement, control, and inclusion shape users' perceptions of ownership~\cite{dhillon2024shaping,hwang202580}. 
Our findings further highlight the importance of the \textit{sequence} and \textit{role} of those contributions. 
Authorship was better preserved when the user remained the source of communicative intent and retained decision authority over how that intent was expressed. 
Features such as localized insertions and replacements supported this process because participants could evaluate AI suggestions against something they had already written rather than adopting an independently produced response.

This distinction becomes particularly consequential in peer support because authorship has implications for both sides of the interaction. 
For the peer supporter, it shapes whether the final response feels like something they genuinely meant and can stand behind. 
For the support seeker, authorship can shape whether the response is interpreted as evidence that another person personally engaged with their disclosure. 
Thus, two responses with similar wording may carry different social meanings depending on how they were produced.
More broadly, our findings suggest that ``AI-assisted'' communication should not be treated as a single category. 
Generating an entire response, revising a user-written response, suggesting one localized change, and checking an otherwise human-written response represent meaningfully different forms of human--AI collaboration. 
Understanding authorship in such systems therefore requires attending not only to the final artifact, but also to the process through which the person and AI arrived at it.

\subsection{From Individual Assistance to Community-Level Trust}

A third implication of our findings is that the consequences of AI-assisted peer support extend beyond the person using the writing tool and the individual response they produce. 
At the level of a single interaction, a peer supporter may find AI assistance useful, retain control over the response, and produce text that is supportive and appropriate. 
Participants nevertheless anticipated problems that would emerge if the same assistance became commonplace across an online community. 
They raised concerns about increasingly similar responses, unreviewed or spam-like comments, and uncertainty about whether responders had actually read and engaged with the posts to which they replied.

Most importantly, participants anticipated that widespread AI assistance could change how \textit{all} supportive comments are interpreted: we found \textit{authorship spillover} (\autoref{sec:trust-community}), in terms of how authorship can be unsettled for every comment. 
The potential harm therefore does not require an AI-generated response to be poor, unsafe, or even correctly identified as AI-assisted. 
This extends concerns in AI-mediated communication beyond evaluations of an individual message or communicator. 
In online peer-support communities, trust also depends on shared expectations about how members participate and what a response represents. 
If those expectations become uncertain, AI assistance used by some members may influence how contributions from other members---including entirely human-written responses---are interpreted. 
% AI writing assistance can therefore create 
This creates a community-level effect that cannot be evaluated solely through individual usability, response quality, or user preference.

This suggests that introducing AI writing assistance into peer-support communities should be understood not only as adding a feature for individual responders, but also as intervening in an existing social ecosystem. 
Different communities may reach different judgments about what forms of AI assistance are consistent with their norms of participation. 
Our own recruitment experience reinforces this point, as some communities and individuals were skeptical of AI-mediated peer support despite the potential benefits identified by participants in our study. 
Questions of disclosure, acceptable levels of assistance, and collective governance therefore follow from the community-level consequences of AI use rather than being only matters of individual interface preference.

\subsection{Design Implications}

% \subsubsection{Assistance That Scaffolds Rather Than Supplies}
\subsubsection{AI Assistance that Scaffolds rather than Supplies Peer Support}

Our findings suggest that future systems should vary the level of AI assistance according to where a peer supporter is struggling, rather than treating complete response generation as the default. 
A responder who does not know how to begin may benefit from an outline or possible response structure; someone who knows what they want to communicate but has difficulty articulating it may need example phrases; and someone who has already written a response may benefit most from localized revision. 
Full response generation can remain available when explicitly requested, but should represent one end of a spectrum of assistance rather than the starting point.

This approach aligns with prior work showing that the amount, timing, and positioning of AI assistance shape users' engagement with the task and experiences of control and ownership~\cite{dhillon2024shaping,hwang202580,xu2025productive}.
At the same time, longitudinal work cautions that repeated AI scaffolding can also foster dependency for some users~\cite{myung2026scaffolding}.
Graduated scaffolding can reduce the linguistic and cognitive burden of providing support while preserving space for responders to determine what they want to say, incorporate their own experiences, and shape how they communicate care.
The design of future systems can additionally make assistance progressive: beginning with minimal support and allowing users to request more only when needed. 
Such a design can also potentially reduce over-reliance on fully generated supportive responses, which participants worried could shift peer support toward increasingly templated interactions.

\subsubsection{Design Revision as In-Situ Guidance}

Our findings also suggest that AI writing assistance can serve a role beyond improving the response currently being written. 
Participants described learning new phrases, noticing considerations they had overlooked, and seeing alternative ways of approaching a supportive response. 
This is particularly relevant for OMHCs, where peer supporters often participate without formal training in supportive communication and may receive little feedback on the quality of their responses.

However, our findings also qualify this opportunity. 
\ppen{} made revisions visible by showing localized changes, but visibility did not necessarily explain \textit{why} a particular revision was preferable. 
Prior work similarly shows that greater understanding of an AI writing assistant does not necessarily translate into better oversight of its suggestions~\cite{rismani2026use}.
A writing assistant that guides supportive communication should expose not only \textit{what} it proposes changing, but also the supportive principle motivating that change. 
For example, a suggestion could indicate that a revision adds emotional validation, softens overly directive advice, invites the poster to elaborate, or introduces actionable guidance. 
Such explanations can remain lightweight and optional so that they do not interrupt the writing process.

This reframes revision assistance as a form of situated guidance: instruction appears against the user's own writing at the moment when it is relevant, rather than as abstract training separated from the act of providing support. 
Prior work has similarly explored LLM-based feedback and practice for developing interpersonal communication skills~\cite{shaikh2024rehearsal,louie2026can}.
Future research can evaluate such a design affordance through longitudinal methods examining whether repeated exposure to such guidance transfers beyond the immediate interaction---for example, whether responders begin incorporating supportive strategies into responses they subsequently write without assistance.

More broadly, AI writing support in this context could be framed less as an \textit{assistant} that produces responses and more as a \textit{writing tutor} that helps users articulate and evaluate their own responses. 
Beyond generation and revision, future systems could offer a ``check'' or ``review'' function that evaluates a user-written response for potential concerns---such as whether it is overly directive, lacks emotional validation, or contains potentially harmful advice---without rewriting the response itself. 
Although such a feature was not examined in our study, it represents an important direction for preserving user authorship while still providing guidance on the quality and appropriateness of supportive communication.

\subsubsection{Supporting User Steering without Ceding Authorship}

% We intentionally designed \ppen{} to move away from a conventional conversational AI interface.
We intentionally designed \ppen{} to move away from a conventional conversational chat-like interaction in which users describe what they want and receive a complete AI-generated response in return.
Instead, our design encourages the user to write their own response, and the IF operates directly on the user's writing and presents localized revisions for consideration. 
This design was motivated by concerns that highly automated assistance can shift users toward evaluating AI output rather than independently engaging with the task~\cite{buccinca2021trust}, while greater AI involvement in co-writing can reduce writers' sense of ownership~\cite{dhillon2024shaping,gmeiner2025intent}.

Our findings, however, reveal a limitation of removing conversational input altogether. 
P07 and P10 wanted a way to communicate what they were trying to say before the system generated or revised their response. 
Without such a mechanism, users could reject or edit an unsatisfactory suggestion, but had limited ability to steer the system toward their intent beforehand. 
Future systems could support \textit{constrained steering}, allowing users to specify their intended message, tone, or whether they want to incorporate a personal experience before requesting assistance~\cite{gmeiner2025intent,lam2026just}. 
This preserves the benefits of communicating intent to a generative model without positioning the AI as the primary author.

\subsubsection{Personalizing Assistance while Preserving Granular User Control}

Building on our findings, we note that future systems can give users greater control over both \textit{how} AI writes and \textit{where} it intervenes. 
Participants wanted to select only the portions of a response where they needed assistance, leaving text they were already satisfied with unchanged. 
They also wanted to compare and return to previous suggestions rather than having successive revisions gradually replace their original writing. 
Therefore, localized and reversible revisions can be a mechanism to help preserve the boundaries between what the user wrote and where AI contributed.

Participants also suggested adapting assistance to their individual writing styles over repeated use. 
Future systems can condition suggestions on a user's prior writing or allow users to explicitly specify aspects of their preferred tone and register. 
However, personalization based on prior communication introduces additional privacy considerations, particularly in sensitive mental health contexts~\cite{yoo2026ai,qin2026ai}. 
Systems should therefore give users control over whether prior writing is used for personalization, what information is retained, and when such personalization is applied. 
More broadly, evaluation of AI-assisted peer support should consider not only whether a response is supportive in the abstract, but also whether it remains recognizable and acceptable as something the peer supporter themselves would say.

\subsubsection{Making AI Involvement Transparent and Community-Governed}

Our findings on AI-authorship and community-level concerns raise a corresponding design question around how AI involvement should be made visible.
A simple binary label such as ``AI-generated,'' however, may poorly represent how systems such as \ppen{} are actually used. 
A peer supporter who writes an entire response and accepts one suggested phrase has engaged with AI differently from someone who submits a fully generated response with no modification. 
Treating these cases identically could obscure rather than clarify authorship.
Therefore, future platform designs can maintain more granular provenance about how assistance contributed to a response. 
For example, they might distinguish between responses that were \textit{AI-generated}, \textit{AI-revised}, or \textit{lightly AI-assisted}, rather than representing AI involvement as binary. 
We do not argue that these categories should necessarily be displayed in the same way across all communities. 
Instead, provenance can provide the underlying information from which communities develop appropriate disclosure practices.
Likewise, platforms can provide controls to the support seeker to indicate whether or not they would like to receive responses co-written with AI assistance, similar to existing audience or privacy controls on online platforms.

This also exposes a tension between transparency and participation, consistent with prior work identifying tradeoffs between transparent attribution and users' experiences of agency in AI co-writing~\cite{hu2026plotania}. 
Disclosure allows support seekers to interpret the origin of what they receive, but disclosure itself may lower perceived authenticity even when the resulting response is supportive. 
Conversely, concealing AI involvement may preserve immediate acceptance while contributing to the broader uncertainty about authorship that our participants described. 
Future work should therefore examine not simply whether AI use should be disclosed, but what level and form of disclosure allows recipients to make meaningful distinctions among different forms of human--AI co-writing.

At the same time, how these forms of assistance are governed need not be uniform across communities.
Instead of imposing a universal platform-wide policy, platforms can allow communities and moderators to participate in determining whether and how AI assistance is available. 
Community-level configurations, for example, on platforms such as Reddit, may determine the level of AI assistance that is permitted, e.g., full generation, revision-based assistance, and whether assistance is restricted for particular categories of high-risk posts, how AI involvement is disclosed, and what community data can be used to personalize the system. 
Such configurations can make AI assistance part of existing community governance, where the role of AI in peer support can remain human- and community-in-the-loop.

\subsection{Ethical Implications}

AI-assisted peer support introduces an asymmetry between those using the assistance and those receiving its outputs. 
The peer supporter may benefit from reduced effort and greater confidence, while the support seeker may experience the same assistance as diminishing the perceived human attention behind the response. 
Because support seekers do not necessarily choose whether the responses they receive are AI-assisted, deployment should consider whose interests automation serves and whose experience it alters.

A second ethical consideration concerns collective consent. One member's willingness to use AI assistance does not imply that others have agreed to an environment where supportive responses may be AI-mediated, which matters especially in communities organized around expectations of peer-to-peer human support.
Platforms should therefore involve moderators and members in decisions around whether assistance is enabled, what forms are appropriate, and what is expected around disclosure and provenance~\cite{peng2020exploring,wang2025practice}.

Finally, integrating AI assistance within an online mental health platform may provide greater control over how sensitive content is transmitted and processed than requiring users to copy posts into external systems. 
However, integration alone
does not guarantee privacy: any deployment still requires explicit policies on what data reach models, whether interactions are retained, whether prior writing is used for personalization, and who can access these data~\cite{zhong2025considerations}---particularly where posts contain highly sensitive disclosures.
Accordingly, we caution against misinterpreting our study as an umbrella endorsement of AI systems in mental health contexts.  
 %2 page
\subsection{Limitations and Future Directions}\label{sec:limitations}
Our study has limitations which also suggest interesting future directions. 
First, our small sample and single one-hour session preclude generalizable or long-term effectiveness claims, and a number of participants were not active support providers---though this positioned them well to speak to the barriers assistance is meant to lower. Second, a critical point emerged during recruitment: when we circulated the study among moderators and within OMHCs, multiple individuals expressed criticism and skepticism about the use of AI to write mental health supportive responses. 
We clarified that our goal was not to advocate for such a tool but to critically examine and understand its usefulness, if any, along with its limitations and concerns; some appreciated this framing and participated, while others remained skeptical and declined. 
Self-selection therefore plays an important role here: our participants were at least open to AI-assisted writing for peer support, so our findings may not capture the perspectives of those holding stronger negative views. Future work should examine these perspectives directly, including community norms around the acceptability of AI for peer support, how such use affects perceptions of authenticity and trust, and what communities expect around disclosure, consent, or platform-level governance.

Further, our participants resided in the U.S. and reported English as their first language, limiting our understanding of how AI-assisted peer-support writing is perceived by people with limited English proficiency and across linguistic and cultural contexts where norms of expressing and receiving support may differ; future work should examine how such systems can support these populations.
Finally, we primarily examined participants acting as support providers. Although we asked them to consider both perspectives during the think-aloud tasks and interviews, these reflections are not direct evidence of how support seekers would feel about receiving---or knowingly
receiving---a response co-written with AI. Future research should therefore pursue two-sided studies of both seekers and providers, including how different forms and degrees of AI assistance are experienced by those receiving the resulting support.

 %2 page
\section{Conclusion}

In this work, we designed and evaluated \ppen{}, an AI-assisted writing system that supported peer supporters in generating and revising responses to posts in online mental health communities. 
Through a user study with 15 participants, we found that AI assistance reduced some of the cognitive and emotional burden involved in composing supportive responses and helped participants identify alternative ways of communicating support. 
At the same time, participants preferred assistance that preserved their role in determining what to say and how to say it, and they raised concerns about how AI involvement could affect perceived authenticity, authorship, and trust among peer supporters, support seekers, and the broader community. 
Our findings showed that the value of AI-assisted peer support could not be understood through response quality alone, because the process through which a response was written also shaped what that response meant as an act of human support. 
This study highlights the importance of designing AI writing assistance that scaffolded rather than replaced supportive communication, while preserving user agency and accounting for the norms and expectations of the communities in which such tools were situated. %2 page


\begin{acks}
% \section*{Acknowledgments}
This work was supported in part by the Jump ARCHES endowment through the Health Care Engineering Systems Center at the University of Illinois and the OSF Foundation.

\end{acks}

\bibliographystyle{ACM-Reference-Format}
\bibliography{0paper}

%
% TC:ignore
\appendix
\setcounter{table}{0}
\renewcommand{\thetable}{A\arabic{table}}

% \clearpage
\section{Appendix}
\label{app:findings}

\begin{table*}[h]
\centering
\sffamily
\footnotesize
\renewcommand{\arraystretch}{0.65}
\caption{Summary of participant-reported strengths, gaps, and resulting design implications, organized by design goal and by concerns surrounding community trust.}
\setlength{\tabcolsep}{2pt}
\resizebox{\columnwidth}{!}{\begin{tabular}{p{0.28\columnwidth}p{0.4\columnwidth}p{0.31\columnwidth}}
\textbf{\ppen{}: Perceived Strengths} & \textbf{\ppen{}: Perceived Gaps} & \textbf{Design Implications} \\
\toprule

\rowcollight \multicolumn{3}{c}{\textbf{DG1: Lowering the Burden of Composition}}\\
$\bullet$ Provides a launch pad, easing both the search for words and the emotional weight of engaging with distress (P01, P05) \newline
$\bullet$ Faster and less draining than unassisted writing, including in high-stakes moments when responders panic (P02, P13) \newline
$\bullet$ Raises willingness and confidence to reply to emotionally heavy posts (P01)
&
$\bullet$ IF is slow, repetitive, and lengthy, interrupting composition (P04, P11) \newline
$\bullet$ Text is at times generic or clich\'ed, with occasional spelling and grammar errors (P01, P08, P13) \newline
$\bullet$ Output far from the user's intent or voice makes editing costlier than writing (P08)
&
\textbf{Scaffold rather than supply} \newline
$\bullet$ Vary assistance by where the user is struggling: an outline or response structure to begin, example phrasings to articulate, localized revision to refine \newline
$\bullet$ Make support progressive, starting minimal and expanding on request, with full generation as one end of a spectrum rather than the default \newline
$\bullet$ Preserve room for users to decide what to say and to draw on their own experiences, guarding against over-reliance on generated text
\\
\hdashline

\rowcollight \multicolumn{3}{c}{\textbf{DG2: Iterative Collaboration}}\\
$\bullet$ Easy-to-edit output leaves room for personalization \newline
$\bullet$ Insert/replace controls allow selective adoption, experienced as ``50--50'' collaboration (P10, P12, P13) \newline
$\bullet$ Seeing the original beside suggestions supported deliberate choice among options (P07) \newline
$\bullet$ Ownership felt strongest when writing first and revising after (P12)
&
$\bullet$ Can feel like proofreading AI output rather than collaborating (P05, P14) \newline
$\bullet$ Text the user did not write feels inauthentic regardless of its quality (P08) \newline
$\bullet$ Suggestions miss the user's voice and cannot draw on personal experience (P08, P13) \newline
$\bullet$ Intent can only be corrected after a suggestion arrives, not stated beforehand (P07, P09, P10)
&
\textbf{Steering without ceding authorship} \newline
$\bullet$ Let users specify intended message, tone, or whether to include a personal experience before assistance is produced \newline
\textbf{Granular and reversible control} \newline
$\bullet$ Let users target only the portions where they want help, leaving satisfactory text untouched \newline
$\bullet$ Keep revisions reversible and comparable, so the boundary between user and AI text stays visible \newline
$\bullet$ Adapt to the user's style over repeated use, with user control over whether prior writing is retained and applied
\\
\hdashline

\rowcollight \multicolumn{3}{c}{\textbf{DG3: Guiding Supportive Writing}}\\
$\bullet$ Surfaces phrasings, questions, and considerations users would not produce alone (P02, P03, P06, P12, P13, P15) \newline
$\bullet$ Offers more concrete and action-oriented advice than users' own writing (P02) \newline
$\bullet$ Exposes alternative approaches rather than only refinements, with some reporting lasting change in how they respond (P01, P06)
&
$\bullet$ Shows what a change is, but not why it is more supportive (P09) \newline
$\bullet$ Red/green markup is hard to read, and the targeted span hard to locate in a longer draft (P05, P09, P11, P14, P15) \newline
$\bullet$ Lengthy suggestions blur substantive guidance and stylistic rewriting
&
\textbf{Revision as in-situ guidance} \newline
$\bullet$ State the supportive principle behind each suggestion, e.g.\ adding validation, softening directive advice, inviting the poster to elaborate, or introducing actionable guidance \newline
$\bullet$ Keep such explanations lightweight and optional so they do not interrupt writing \newline
$\bullet$ Add a review mode that flags concerns in a user-written response without rewriting it \newline
$\bullet$ Frame assistance as a writing tutor, and examine longitudinally whether guidance transfers to unassisted writing
\\
\hdashline

\rowcollight \multicolumn{3}{c}{\textbf{DG4: Platform Integration}}\\
$\bullet$ Intuitive and familiar; assistance blends into Reddit and suits regular use (P01, P03, P04, P10) \newline
$\bullet$ Output matches the tone and norms of Reddit comments (P08) \newline
$\bullet$ Support for direct integration into Reddit, including from a moderator (P03, P04, P10)
&
$\bullet$ Unmarked cursor-following insertion leaves placement unpredictable and verifiable only after the fact (P01, P02, P05, P12, P15) \newline
$\bullet$ Output sometimes reads as formal or AI-like rather than community-native
&
\textbf{Legible and configurable in place} \newline
$\bullet$ Mark insertion points explicitly, e.g.\ a blinking cursor or an in-place preview (P05, P12, P15) \newline
$\bullet$ Let communities and moderators configure the assistance available in their space rather than imposing a uniform platform-wide policy
\\
\hdashline

\rowcollight \multicolumn{3}{c}{\textbf{Open Challenges and Community Trust}}\\
$\bullet$ Acceptable for responding to anonymous users, but not to known individuals (P01)
&
\textbf{Perspective of Peer Supporters} \newline
$\bullet$ Discomfort using AI on suicidal or emotionally heavy posts regardless of quality; AI should not replace what humans do naturally (P05, P08, P09, P11, P13) \newline
$\bullet$ May detach responders and erode empathy by removing the need to think before commenting (P01, P08) \newline

\textbf{Perspective of Support Seekers} \newline
$\bullet$ AI authorship read as less serious or as a breach of trust (P01, P08) \newline
$\bullet$ Seekers want human support and may feel worse if the effort was outsourced: the same words carry a different meaning (P04, P09, P10) \newline

\textbf{Perspective of Broader Community} \newline
$\bullet$ Homogenized, artificial responses, and scope for spam or unreviewed posting (P11, P13, P14) \newline
$\bullet$ Awareness that such tools exist casts doubt on all comments, including human-written ones (P01, P02)
&
\textbf{Transparency and provenance} \newline
$\bullet$ Record graded provenance---generated, revised, or lightly assisted---rather than a binary AI label \newline
$\bullet$ Treat provenance as the basis from which communities develop disclosure practices, weighing transparency against the authenticity that disclosure itself may cost \newline
$\bullet$ Give support seekers controls over whether they receive AI co-written responses \newline

\textbf{Community-governed norms} \newline
$\bullet$ Let communities set the permitted level of assistance, restrictions for high-risk posts, how involvement is disclosed, and what data may personalize the system
\\
\bottomrule
\end{tabular}}
\label{tab:findings}
\Description{A three-column table summarizing participant-reported strengths, gaps, and design implications for
PeerPen. The columns report perceived strengths, perceived gaps, and the design implications derived from them.
Full-width section rows divide the table under the four design goals---Lowering the Burden of Composition,
Iterative Collaboration, Guiding Supportive Writing, and Platform Integration---followed by a final section on
open challenges and community trust. Entries are bulleted, with participant IDs indicating who raised each point.
Implications are grouped under headings corresponding to our design implications subsections, and the final
section groups trust concerns by the perspective of peer supporters, support seekers, and the broader community.}
\end{table*}

\end{document}
% TC:endignore